\pdfoutput=1
\documentclass[manuscript]{acmart}

\AtBeginDocument{%
  \providecommand\BibTeX{{%
    \normalfont B\kern-0.5em{\scshape i\kern-0.25em b}\kern-0.8em\TeX}}}

\usepackage{xcolor}

\usepackage{soul}
\usepackage{enumitem,xcolor}
\newcommand{\revisions}[1]{{\textcolor{black}{#1}}}
\setstcolor{red}
\newcommand\redsout{\bgroup\markoverwith{\textcolor{red}{\rule[0.5ex]{2pt}{0.4pt}}}\ULon}
\usepackage{balance}
\usepackage{csquotes}
\setcopyright{acmcopyright}\acmConference[Preprint]{Author Preprint}

\begin{document}

\title[Human-AI Guidelines in Practice]{Human-AI Guidelines in Practice: Leaky Abstractions as an Enabler in Collaborative Software Teams}


\author{Hariharan Subramonyam}
 \affiliation{%
  \institution{Stanford University}
   \country{USA}
 }
 \email{harihars@stanford.edu}

 \author{Jane Im}
 \affiliation{%
  \institution{University of Michigan}
   \country{USA}
 }
 \email{imjane@umich.edu}
 
 \author{Colleen Seifert}
 \affiliation{%
 \institution{University of Michigan}
   \country{USA}
 }
 \email{seifert@umich.edu}
 
 \author{Eytan Adar}
 \affiliation{%
   \institution{University of Michigan}
   \country{USA}
 }
 \email{eadar@umich.edu}

\renewcommand{\shortauthors}{Subramonyam et al.}

\begin{abstract}

\revisions{In conventional software development, user experience (UX) designers and engineers collaborate through separation of concerns (SoC): designers create human interface specifications, and engineers build to those specifications. However, we argue that Human-AI systems thwart SoC because human needs must shape the design of the AI interface, the underlying AI sub-components, and training data. How do designers and engineers currently collaborate on AI and UX design? To find out, we interviewed 21 industry professionals (UX researchers, AI engineers, data scientists, and managers) across 14 organizations about their collaborative work practices and associated challenges. We find that hidden information encapsulated by SoC challenges collaboration across design and engineering concerns. Practitioners describe inventing ad-hoc representations exposing low-level design and implementation details (which we characterize as \textit{leaky abstractions}) to ``puncture'' SoC and share information across expertise boundaries. We identify how leaky abstractions are employed to collaborate at the AI-UX boundary and formalize a process of creating and using leaky abstractions.}

\end{abstract}

\ccsdesc[500]{Human-centered computing~HCI theory, concepts and models}
\ccsdesc[500]{Software and its engineering~Designing software}
\ccsdesc[500]{Software and its engineering~Abstraction, modeling and modularity}
\ccsdesc[500]{Software and its engineering~Collaboration in software development}
\keywords{Human-AI systems, UX design, AI applications, Industry practices, design processes}

\maketitle

\section{Introduction}
\revisions{In conventional software development, work processes for user experience (UX) designers and software engineers are optimized for efficiency through \textit{separation of concerns} (SoC)~\cite{bardini2000bootstrapping,grudin2017tool, seffah2005human}. UX roles focus on human psychology and design by working with end-users to define system requirements. Software engineers skilled in computer programming then implement those requirements in a system~\cite{seffah2005human}. For example, to create a conventional (non-AI) \textit{To-Do List} application, the designer first gathers information from different end-users (students, IT professionals, educators, etc.) on how they define and manage their tasks. Based on those insights, the designer generates several interface alternatives to find one that meets end-user needs. Using knowledge about graphical user interfaces (GUI), established design guidelines, and design tools, designers generate specifications for \textit{all} aspects of software behavior, including UI designs, functional requirements, style guides, data requirements such as task description lengths, interaction maps, and evaluation criteria~\cite{maguire2002user}. Finally, this highly controlled and abstracted knowledge is handed to engineers through serial coordination~\cite{seffah2005human} to be translated into technical requirements and subsequently implemented as software code~\cite{ patel2010lowering}.}

\revisions{However, the efficiency of SoC that works well in conventional software development may fail for Human-Centered AI (HAI) systems. HAI systems differ in several important ways: (1) they offers greater capacity for human-like intelligent behavior, (2) they dynamically learn and adapt their behavior over time through feedback and learning, and (3) their outputs can be non-deterministic, making it difficult to align output presentation with end-user expectations~\cite{yang2020re}. By examining the complex dependencies across different components in the AI lifecycle, prior research has laid out desiderata for AI systems. This includes ensuring contextually relevant datasets and comprehensive and comprehensible verification of AI models~\cite{amershi2019software, ashmore2021assuring}, adapting for AI uncertainties and failures in human interface design~\cite{amershi2014power, amershi2019guidelines, PAIR}, and incorporating human and social interpretations for AI model design~\cite{baumer2017toward}. These demands make it challenging to separate current UX work processes from AI software development tasks. Consider designing a ``smart'' \textit{To-Do List} application to automatically create task items from email content (e.g.,~\cite{peopleai}). In taking a human-centered approach, UX roles must identify representative AI dataset characteristics based on diverse users covering a range of expressive email tasks. They need to support creating ``ground truth'' data to define how users want to generate tasks from those emails. UX designers need to provide inputs about AI model behavior by considering how the AI experience will integrate with end-user task workflow: what to automate, when to offer assistance, and when to maintain human control of tasks. Finally, designers must consider uncertainties in AI model outputs and design interface adaptations for explainability, failures, feedback, and hand-off. Consequently, combining these rationalistic and design goals for HAI requires multidisciplinary collaboration~\cite{auernhammer2020human,subramonyamProcessModel2021}.}

\revisions{While a growing body of HAI design guidelines point towards blending AI and UX work practices~\cite{amershi2019guidelines, PAIR, Appleguidelines}, we still lack concrete knowledge on \textit{how} to achieve such collaboration. Recent work has highlighted numerous concerns due to SoC at the AI-UX boundaries, including challenges in understating AI capabilities~\cite{dove2017ux,yang2020re}, difficulty in specifying AI system requirements~\cite{yang2017role}, and prototyping HAI interfaces~\cite{yang2019sketching}. Further, current practices in which the AI components are developed before envisioning the human user experience (i.e., AI-first design process) have led to AI systems that do not align with human needs (e.g., incorrect labeling~\cite{Facebook}, biased auto-cropping~\cite{twitterauto} of social media photos, faulty facial recognition features~\cite{buolamwini2018gender}, etc.). Understanding how industry practitioners can and should collaborate across technical and non-technical roles is essential for producing HAI systems that can be successful in real-world settings. In this work, our goal is to improve our understanding of how industry practitioners (both UX and AI roles) work and the challenges and solutions they have identified in creating \emph{human-centered} AI systems. Ultimately, we aim to propose a better approach for team-based HAI development based on the derived insights.}

\begin{displayquote}
\textit{\revisions{\textbf{Research Question 1:} What challenges do HAI designers and engineers face in creating HAI systems following the standard SoC process?}}

\textit{\revisions{\textbf{Research Question 2:} How do designers and engineers adapt their work processes to improve HAI outcomes?}}

\textit{\revisions{\textbf{Research Question 3:} How might HAI teams integrate concerns across disciplinary boundaries to align human and AI needs?}}
\end{displayquote}

\revisions{To investigate these questions, we first collected and analyzed a total of 280 HAI design guidelines across different industry sources. From this analysis, we derived a component model for designing HAI applications that span data, ML model, user interface, and end-user mental models (see Figure~\ref{fig:model}). Using our model as a guide, we interviewed 21 industry practitioners (UX designers, AI engineers, data scientists, and product managers) across 14 different organizations to understand their current practices for creating HAI systems. Through the interviews, we identify sources of friction in the HAI development process and uncover how practitioners currently bridge the design-engineering boundary. Our findings show that current HAI workflows rarely begin with end-user needs due to the challenges for designers in defining AI experiences upfront. In practice, HAI designers are now working to shape user experiences around novel AI capabilities. However, successful teams circumvent collaboration challenges by delaying commitment to solutions until later in the design process. Finally, we highlight specific practices around knowledge sharing across expertise boundaries that oppose established SoC practices. As opposed to SoC and information hiding, we find that in successful teams, designers and engineers communicate across boundaries through ``leaky'' abstractions that facilitate a collaborative design process. Many existing practices have evolved in a haphazard fashion. We attempt to better formalize the use of leaky abstractions.}

\revisions{We contribute to the current understanding of challenges faced by UX roles~\cite{dove2017ux, yang2019sketching,yang2020re} and AI roles~\cite{zhang2020data, muller2019data, cai2019software} in developing AI systems that align with human needs,  values and are useful and usable by people~\cite{ gabriel2020artificial, manning2020, mohamed2020decolonial,riedl2019human, xu2019toward, shneiderman2020human}. Through the lens of the component model of HAI guidelines, we describe the limitations of existing SoC practices that are optimized for efficiency but hinder cross-disciplinary collaboration. Further, we discuss alternatives to standard software development workflows to bridge knowledge boundaries and highlight solutions for collaboration and co-design of HAI systems. Finally, through our discussion, we offer advice for software organizations to realize HAI guidelines and make recommendations for HAI pedagogy. }
\section{Related Work}
\revisions{Human-Centered AI frames AI as technology that \textit{``augments the abilities of, addresses the societal needs of, and draws inspiration from human beings''}~\cite{manning2020}. Based on this vision, research in HCI and AI communities has characterized and detailed domain-specific viewpoints~\cite{amershi2014power,buolamwini2018gender,dignum2017responsible,  stark2019facial}, identified challenges~\cite{dove2017ux,lwakatare2019taxonomy,yang2020re}, and put forth requirements and strategies~\cite{ amershi2019guidelines, ashmore2021assuring,baumer2017toward} to operationalize HAI. Here we synthesize what is known about current human-centered software development (HCSD) processes, expertise, design workflows, and boundary representations to identify challenges to designing HAI systems. Through this synthesis, we highlight the gap we aim to address.}

\subsection{\revisions{Human-Centered Approaches in Industry Software Teams}}

\subsubsection{\revisions{Modular Software Development:}} \revisions{HCSD is a complex problem requiring knowledge and expertise beyond what any single person can possess. When multiple individuals are involved (UX designers, software engineers, and database experts, etc.), the preferred approach is to decompose the system into modules and tasks that can be carried out relatively independently by different people~\cite{agre1997computation,seffah2005human}. Often, system modules and work-team structures observe a homomorphic relation~\cite{conway1968committees}. For instance, UX professionals create the user interface, and engineers implement the underlying functionality. Specific to HAI, Amershi et al. propose a nine stage software engineering workflow for machine learning that begins with specifying model requirements and subsequently, data collection, features engineering, and model development~\cite{amershi2019software}. Prior studies with data scientists~\cite{muller2019data, muller2019human, zhang2020data} have uncovered numerous challenges to realize such workflows, including involvement of non-technical roles in technical work stages (features engineering, model development), difficulty in deriving features based on deep domain knowledge, and data collection and labeling. On the other hand, in assuming a \textit{material design} approach to HAI, Yang et al. study UX practitioners and their design processes for HAI. Through this investigation they highlight challenges for realizing the double diamond UX process model for AI interface design~\cite{council2005double,yang2018machine, yangre}. Further, while agile methodologies have improved HCSD workflows in conventional software~\cite{przybilla2018combining,lucena2016ibm}, the short development cycles and rapid turnarounds are infeasible for AI development which requires a longer time to design and implement~\cite{yang2018investigating}. }

\subsubsection{\revisions{Information Hiding, Abstractions, and Collaboration:}} \revisions{In multidisciplinary teams, to reduce dependencies between tasks, team members first define the module's outward-facing \textit{interface} while the \textit{implementation} details are abstracted from one another (i.e., information hiding)~\cite{parnas1972criteria,henning2007api}. In HCSD, designers take a ``UX first'' approach to design the system's `user interface' ~\cite{jaredAPI}. Here, the user interface can be considered the highest level module for end-users to invoke. Designers map end-user needs into interface design specifications. Engineers who also understand the language of the user interface can translate interface representation into implementation~\cite{seffah2005human}. In other words, the user interface acts as a natural `seam' for designers and engineers to coordinate. However, such interface-level abstractions quickly break down when designing AI-powered applications. For instance, in investigating how designers sketch experiences for natural language processing (NLP), Yang et al. highlight the challenges to design abstractly and propose the need for ML specific abstractions (e.g., language, capabilities, and experiential qualities of NLP) to support designers~\cite{yang2019sketching,yang2018investigating}. Yet, other work has shown that it can be challenging to enforce strict abstractions~\cite{sculley2015hidden}. In fact, ML is beneficial in cases in which behavior cannot be explicitly specified through software logic~\cite{patel2010lowering,de2004sometimes}. Further, in the case of HAI, the \textit{contract} nature of abstractions hides implementation details that are necessary for designing AI adaptations, such as explainability and feedback~\cite{amershi2019guidelines,cataldo2010architecting}. With AI, designers and engineers need to bridge abstraction levels along a \textit{part-whole hierarchy} to center people in the design of AI sub-components, and within an \textit{implementation hierarchy} to offer interface adaptations to AI uncertainties~\cite{visser2006cognitive}.}

\revisions{In sum, prior work has uncovered limitations of existing HCSD workflows when it comes to HAI development. However, previous studies tended to focus solely on data scientists~\cite{muller2019data, muller2019human, zhang2020data} or designers~\cite{yang2019sketching,yang2018investigating}. It remains an open question on how to handle abstraction in multidisciplinary collaboration between technical and non-technical roles. Our work aims to address this gap.}

\subsection{\revisions{Key Design and Engineering Challenges for HAI}} \label{HAI-collaboration-challenge}

\subsubsection{\revisions{Challenges for Designers:}} \revisions{Design knowledge for human-AI systems is comprised of (1) understanding task characteristics, including type of goals and data representations, (2) machine learning paradigms, (3) human-AI interactions such as machine teaching, and (4) AI-human interactions such as interpretability~\cite{dellermann2019future}. However, current UX designers are not trained in these aspects of HAI systems. First, UX designers lack the expertise to generate design ideas for incorporating AI in human tasks~\cite{dove2017ux,yang2020re}. As a result, they often misunderstand the capabilities of ML models and propose designs that can be difficult to implement~\cite{kayacik2019identifying}. Second, given that AI takes a long time to build~\cite{yang2018investigating}, rapid prototyping with ML through a ``fail fast, fail often'' approach characteristic of UX design is challenging for HAI~\cite{yang2019sketching}. Moreover, AI requires vertical end-to-end prototyping to identify uncertainties and edge cases and to create UI adaptations~\cite{baca2018merging, beaudouin2009prototyping,  corbettinteractive}. However, black-box views of ML make it difficult for designers to understand, design, and evaluate with AI~\cite{helms2017leaky,helms2018design}. Third, UX processes favor creativity and imagination of desired futures, which contradicts AI's emphasis on specificity and accuracy~\cite{yang2017role}. This introduces friction into the design thinking process for HAI systems. }

\subsubsection{\revisions{Challenges for Engineers:}} \revisions{Similarly, engineers focused on algorithms and techniques fail to consider human perspectives during initial experimentation and AI prototyping processes~\cite{hill2016trials,lwakatare2019taxonomy}. Several aspects of HAI design need to be incorporated throughout AI workflow, including identifying model requirements, data collection and labeling, features engineering, and model training~\cite{amershi2019software, hellerstein2017ground, russell2015research}. But expertise in HCI and involvement in exploring human needs are lacking in engineering training. Engineers who are ML novices were shown to experience breakdowns in early-stage software development due to lack of specialized design schemas, insufficient understanding of the design process, and sub-optimal design solutions~\cite{cai2019software,guindon1987breakdowns}. Consequently, even when designers suggest modifications for better human-centered experience design, model and data changes to the AI may be challenging to execute. In AI techniques such as deep learning, it can be challenging to identify specific functional areas to address human user issues~\cite{arpteg2018software}. Further, by focusing on creating the algorithm, engineers often fail to consider the AI experience as a whole and their involvement in UX design tapers~\cite{dove2017ux,gibson2001knowledge}. AI and UX practitioners can benefit from a symbiotic relationship~\cite{ceconello2019design}. HCI perspectives about the user interface can improve AI through better quality feedback on performance~\cite{schnabel2018improving}. For example, AI output presentation can impact end-users' subjective perception of errors and how they adjust their expectations about AI~\cite{kocielnik2019will}.} 

\revisions{To summarize, prior research has separately uncovered design and engineering challenges and respective knowledge barriers for HAI. However, we lack an understanding of the entire design and engineering pipeline for creating HAI systems in a multidisciplinary team-based approach. In this work, we build on existing research by studying how industry practitioners (both UX and AI roles) collaborate across technical and non-technical roles. This includes challenges that arise in work processes, workarounds the practitioners have created to address the challenges, and their needs for solutions that do not yet exist. We propose a concrete approach for successful team-based HAI development based on this understanding. }

\subsection{Boundary Representations for Collaboration}

\subsubsection{\revisions{Role of Boundary Representations:}} \revisions{In complex domains such as HAI, teams would ideally address knowledge differences or ``symmetry of ignorance'' between HCI and AI professionals through collaboration and social creativity~\cite{fischer2000symmetry}.  Prior work on software collaboration has identified three types of knowledge boundaries, including (1) assembling--how information should be structured, (2) designing--how information artifacts are designed, and (3) intended user interaction--how users interact with designed information~\cite{winkler2015recurrent}. The goal for collaboration is to bridge the knowledge boundaries described in section \ref{HAI-collaboration-challenge} to acquire \textit{common ground} for interaction~\cite{stalnaker2002common}. Common ground in collaborative work includes \textit{content} common ground and \textit{process} common ground~\cite{clark1991grounding, mao2019data}. In HAI, the content common ground is the \textit{data} which forms the backbone of machine learning (AI) applications~\cite{subramonyamProcessModel2021}, and the process entails the design~\cite{yang2020re} and engineering~\cite{amershi2019software} in creating both the AI and the UX. Further, these knowledge boundaries can be bridged by either \textit{converging} content and process knowledge bases through elaboration, discussion, and negotiation of dependencies across boundaries (i.e., traversing knowledge boundaries) or through knowledge \textit{transcendence} by integrating just the necessary information for collaboration through co-created scaffolds (i.e., parallel representations) and dialog around scaffolds~\cite{majchrzak2012transcending}.} 

\subsubsection{\revisions{Boundary Objects:}} \revisions{Boundary objects~\cite{leigh2010not,star1989structure}, such as external representations, play a critical role in bridging knowledge boundaries by supporting information sharing, interpretation, negotiation, and co-design. In collaborative design, these representations also include \textit{epistemic} objects such as artifacts of design-pursuit characterized by incompleteness and \textit{technical} objects including design tools that support the process of design inquiry~\cite{ewenstein2009knowledge}. Further, when the boundaries are blurry and non-standard, material artifacts support the process of characterizing boundaries and collaboration, which are called boundary negotiation artifacts~\cite{lee2007boundary}. These artifacts consist of (1) self-explanation artifacts for learning, recording, organizing, remembering, and reflecting, (2) inclusion artifacts for proposing new concepts, (3) compilation artifacts to coordinate and align knowledge, (4) structuring artifacts to establish principles at the boundaries, and (5) borrowing artifacts that are repurposed in unanticipated ways across communities to augment understanding~\cite{lee2005between}. The eventual representation created by the differing expertise through collaboration is the artifact's \textit{specifications} encapsulating the \textit{what}---the artifact product itself, the \textit{how}---the procedure by which it should be implemented, and the \textit{why} (design rationale)---the reason why the design should be as it is~\cite{visser2006cognitive}. }
 
\revisions{In conventional software development, prototypes are commonly used as boundary objects~\cite{huber2020use}. They serve to bind user needs and technical information and can include design prototyping, prototypes for verification, prototypes for exhibition, etc.~\cite{ hirota2017design,von1994sticky}. The need for boundary objects for AI interface design has been emphasized in recent studies~\cite{yang2018investigating}. But as collaborations for HAI systems still lack standardization, the concept of boundary negotiation artifacts is also likely to be important and relevant. Prototypes for HAI should promote agreement in defining task specifications, communicating states of design, identifying references of central notions, and negotiating weights of criteria and constraints~\cite{visser2006cognitive}. Given the collaboration challenges described in section \ref{HAI-collaboration-challenge}, we need new prototyping approaches for defining specifications that include process, content, structure, and form~\cite{leveson2000intent}. Further, prototypes should embody a means-ends hierarchy for envisioning HAI in which each level specifies the what, the how of the level below, and the why of the level above~\cite{leveson2000intent}. Prior work has identified characteristics of effective boundary prototypes, including interpretive flexibility, plasticity~\cite{kertcher2018boundary}, and translucency~\cite{cataldo2010architecting,ebling1998translucent}. These characteristics support (1) establishing a shared syntax, (2) concrete means to learn about differences and dependencies, and (3) joint knowledge transformation without causing information overload~\cite{carlile2002pragmatic}.} 

\revisions{Our work studies the boundary negotiation artifacts to overcome knowledge barriers and achieve standardization across technical and non-technical roles. We further propose alternative software development workflows to accommodate the practice of boundary negotiation and blending.}

\section{Study 1: Analysis of HAI Design Guidelines for Collaboration Recommendations}

\begin{figure*}[t!]
\centering
\includegraphics[width=\textwidth]{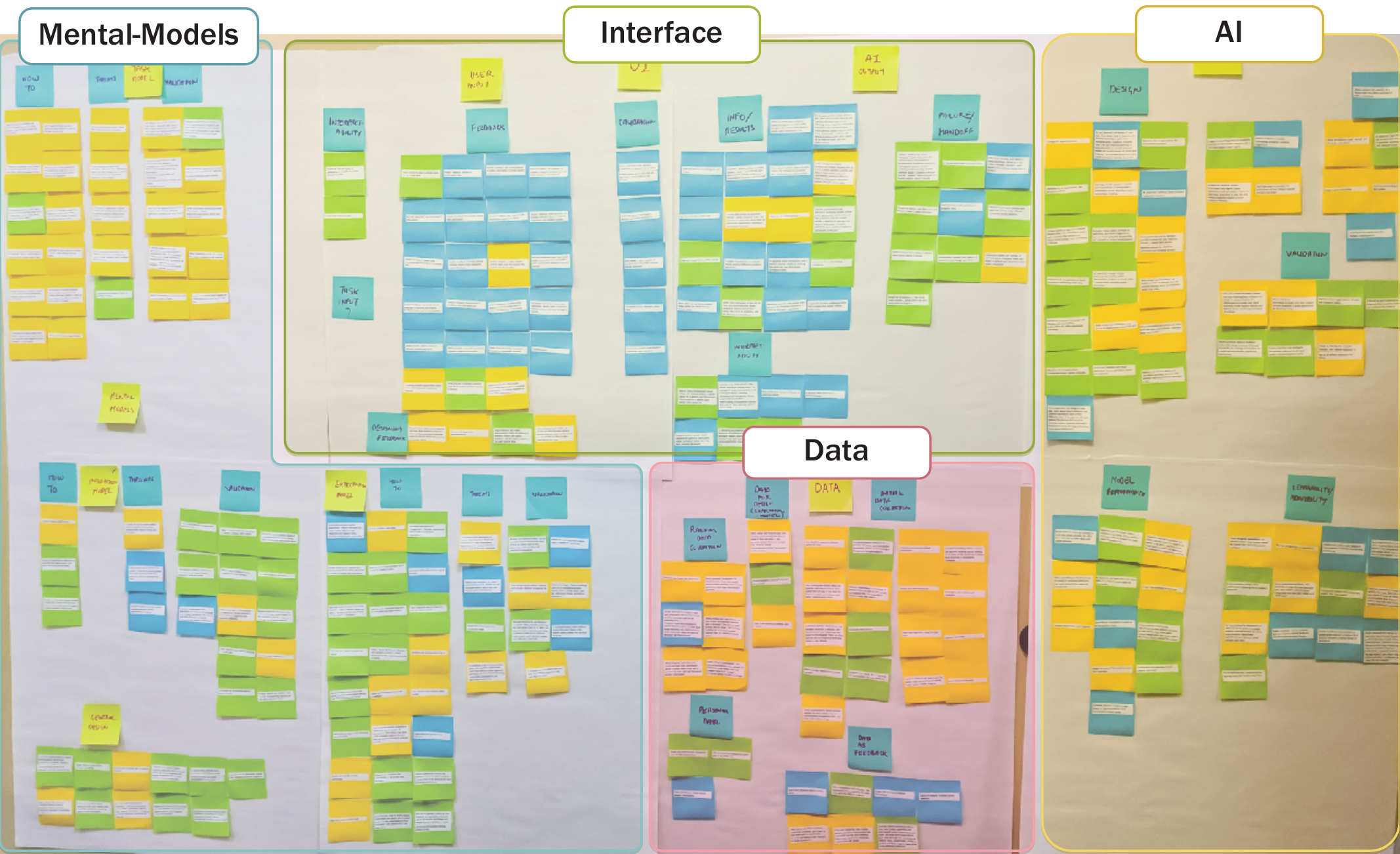}
\caption{Affinity Diagram of HAI Guidelines}
\Description{Picture of the Affinity Wall with the four clusters annotated}
\label{fig:ad}
\end{figure*}

\revisions{To address our research questions on collaborative HAI practices, we began by determining a consensual view of recommended industry design practices for HAI. We collected HAI design guidelines from major industry sources to characterize current understanding of collaboration requirements in the field. Then, we synthesized the recommendations as a set to create a comprehensive model of HAI guidelines. This summary model serves as structure for our interviews (Study 2) with industry practitioners to organize inquiries about actual design processes used in industry projects. }

\subsection{\revisions{Method}}
\subsubsection{\revisions{Data Collection}}
\revisions{Various companies have offered recommendations for human-AI application design based on their internal design and development practices. Our primary sources include Microsoft's ``Guidelines for Human-AI Interactions''~\cite{microsoft, amershi2019guidelines}, Google's ``People + AI Guidebook''~\cite{PAIR}, and Apple's ``Human Interface Guidelines''~\cite{Appleguidelines}. In addition, we collected recommendations published through formal and informal reports by industry practitioners, including ``Human-AI Guidelines Cheat-sheet'' ~\cite{Lovejoy} and ``Deloitte Insights''~\cite{Deloitte}. If a guideline combined multiple recommendations in a single sentence, we split the guideline into individual recommendations. In total, we collected 280 separate design guidelines across these sources. We arrived at a final 249 after removing or combining similar guidelines.}

\subsubsection{\revisions{Analysis}} \revisions{The first author conducted an affinity diagramming exercise~\cite{scupin1997kj} to identify key topic hierarchies in the guidelines (Figure~\ref{fig:ad}). To create the affinity notes, each guideline was printed on paper and pasted onto a physical sticky note. By mounting blank flipchart sheets onto a wall, the first author grouped individual notes based on perceived affinity. The authors discussed the emergent hierarchies of clusters and determined that the HAI guidelines stress the \textit{goal} of combining AI and UX design but do not describe (or prescribe) \textit{how} designers and engineers might collaborate. Based on these clusters, we developed a component model of human-AI design guidelines (Figure~\ref{fig:model}) and a set of questions for structuring the interview. Here, we summarize the guidelines and questions about individual components of the model.}

\subsection{\revisions{Findings:} A Component-Model Representation of HAI Guidelines}
As shown in Figure~\ref{fig:model}, the model consists of four main components, including (1) human mental models, (2) user interface, (3) AI models, and (4) training data. As indicated by the arrows, humans (and their mental model) are tightly linked to all other components to realize human-centered design.

\subsubsection{Human mental-models:} This set of 89 guidelines focuses on understanding end-user needs in order to design AI features. Specifically, they target (1) understanding how end-users would perform a task on their own and the challenges they might face; that is, the \textit{task model}; (2) understanding people's expectations about what the AI should do, and setting expectations for people about AI behavior, which we call the \textit{expectation model}, and (3) identifying the best type of AI interaction experience given the situational context; namely, the \textit{interaction model}. The guidelines suggest that designers and engineers  elicit these human mental models based on their application vision (or context) and develop a shared understanding for downstream AI and UX design choices. For instance, one of the guidelines about the task model recommends identifying opportunities for AI by understanding the existing task workflow: \textit{``mapping the existing workflow for accomplishing a task can be a great way to find opportunities for AI to improve the experience~\cite{PAIR}''}. During this need-finding process, the guidelines also recommend assessing end-user expectations about AI behavior to find the right balance between control and automation when performing tasks. 

To operationalize these guidelines, designers need to understand AI capabilities and limitations, and they need to share information about human tasks workflows with engineers. However, the guidelines do not specify how to do this: How do UX practitioners understand AI capabilities, implementation assumptions, and needs? How do they formulate expectation models with end-users? And how do they gather, synthesize, and communicate their understanding of human tasks with engineering teams? Our interview questions target these concerns.

\subsubsection{User Interface:} This set of 65 user interface guidelines target the software and hardware interface between end-users and AI. The  recommendations center on lowering the gulf between execution and evaluation~\cite{norman1986user} by designing for (1) end-user inputs and AI outputs, (2) explainability, (3) feedback, and (4) failures and hand-offs. For example, these guidelines recommend that when presenting uncertain AI outputs, we should \textit{``prefer diverse options and, when possible, balance the accuracy of a response with the diversity of multiple options~\cite{Appleguidelines}.''} These guidelines also suggest demonstrating to end-users how to get the best results based on their inputs, conveying how end-user actions will impact future AI behavior, and providing easy ways for users to edit, refine, or recover from AI failure. 

\begin{figure*}[t!]
\centering
\includegraphics[width=\textwidth]{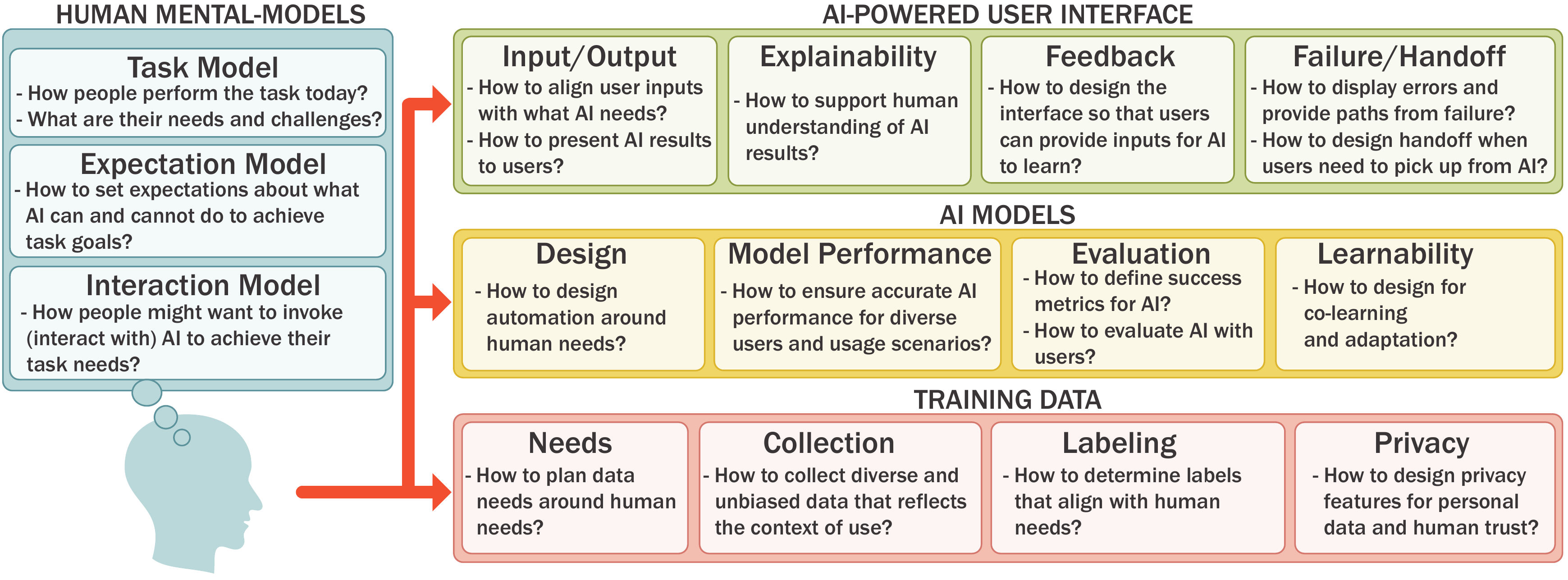}
\caption{Component Model Representation of Human-AI Guidelines}
\Description{Diagram illustrating the four blocks of the component model with topics and questions listed within each block}
\label{fig:model}
\end{figure*}

On their own, UX designers cannot realize these guidelines when making user interface choices. Implementing them requires that UX designers understand low-level implementation details about the AI model. This requires designers and engineers to collaboratively specify (or negotiate) the application programming interface (API) for AI features. Our interview questions addressed API design and the negotiation process between designers and engineers. We asked how designers understand AI failures and collaboratively design how these are experienced by end-users. Existing guidelines do not specify how designers and engineers negotiate about the feedback needed for AI model improvement, or how do teams prototype and assess different interface and API choices.

\subsubsection{AI Model:} These 61 HAI guidelines for AI models focus on designing AI features in a `human-centered' manner. This includes (1) designing the AI based on end-user mental models; (2) designing for co-learning and adaptation; (3) defining model performance in a human-centered way; and (4) evaluating the AI across a range of use scenarios. Regarding the AI model design, guidelines emphasize that the design should reflect information, goals, and constraints that human decision-makers weigh in decisions, avoid unwanted biases and unfair stereotypes, and evaluate the impact of AI ``getting it wrong~\cite{microsoft}.'' AI model guidelines mirror the human mental-models guidelines about the task and expectation model subcomponents of HAI design. Working with these guidelines requires designers to be somewhat knowledgeable about AI implementation choices. Our interview questions thus focus on how engineers communicate about AI assumptions and implementation choices with designers.

Other guidelines recommend defining AI model performance in a human-centered way by considering human values when weighing the cost of false positives and negatives, ensuring that model metrics such as accuracy are appropriate to the context and goals of the overall system, and making conscious trade-offs between precision and recall. For instance, guidelines recommend that \textit{``while all errors are equal to an ML system, not all errors are equal to all people. You will need to make conscious trade-offs between the precision and recall of the system~\cite{PAIR}.''} Similarly, guidelines about evaluating AI features recommend assessing whether model objectives provide a good experience for all users, assessing for safety and whether the AI design performs under the ``realities of the environment in which it to be used.'' We included interview questions to uncover how designers and engineers work together to define model performance metrics and how they evaluate model behavior. 

\subsubsection{Training Data:} According to these 34 guidelines, data needs for training the AI model should reflect human data needs for their tasks. This includes (1) planning data sources, (2) data collection, (3) labeling data, and (4) privacy, security, and ethics of data. For instance, when planning data needs, guidelines recommend aligning them with the task model by asking what information a human will use to perform the task on their own~\cite{PAIR}. For data collection, the guidelines include (1) responsibly sourcing the data, (2) planning data collection to be representative of expected end-users, use cases, and context of use, (3) formatting data in ways that make sense to human users, and (4) collecting only the most essential information from end-users. For labeling data, these guidelines focus on using proper labels; i.e., data labels must reflect the people's diversity and cultural context. 

Implementing these guidelines again requires that designers understand the AI's data needs and the types of computation that AI engineers will apply to the data. Further, they need to work with engineers to define human-friendly data labels, plan data collection, and mitigate problematic biases in the dataset. Our interview questions thus target how teams collaboratively scope data needs based on AI model needs, human task, and expectation models. 

In summary, existing HAI guidelines focus on `what' needs to be done, but they make no recommendations about `how' specific design and engineering processes (user research, data collection, model development, interface evaluation, etc.) serve to align AI development with human-centered design. Nor do they recommend how designers and engineers can bridge their respective knowledge boundaries to acquire a shared understanding to collaborate on HAI design. To answer these questions, we turned to practicing AI and UX professionals in industry to ask about their current processes for HAI design. We structured an initial set of questions using the component model created from affinity clusters and included the concerns highlighted above. To specify the question content, we identified the key \textit{nouns} (e.g., `data', `human-needs') and \textit{verbs} (e.g., `collect', `align') from the guidelines within each cluster. We then translated them into questions about \textit{who} implemented the guidelines, and \textit{how} they did so. For instance, we ask teams about who is involved in collecting data for the AI and how they defined representative data collection needs. As a second example, we ask who is involved in envisioning the AI behavior and how they incorporate human needs into their design. The complete set of interview questions in available in the supplemental material. Our interview study with HAI designers and engineers working in industry aims to identify how they implement these design concerns in their collaborative practice on the job.

\section{Study 2: Interview with HAI Practitioners}

\subsection{\revisions{Method}}

\begin{table*}[t!]
\small
\centering
\begin{tabular*}{0.95\textwidth}{@{}l | l |l| l@{}}
Organization & Interviewee (Years of HAI Experience) & Business Model & Size of Organization
\\\midrule
O1 & S1 (2 yrs) & B2C & 1,000 -- 5,000\\ 
O2 & S2 (3 yrs), S4 (2 yrs), M3 (12 yrs) & B2C & 10,000 -- 50,000\\ 
O3 & M2 (4 yrs), R2 (6.5 yrs), U1 (4 yrs), U5 (3 yrs) & B2C, B2B & $>$ 100,000 \\ 
O4 & M1 (2.5 yrs) & B2B & $<$ 100\\
O5 & D1 (5 yrs) & B2B & $>$ 100,000\\ 
O6 & S5 (3 yrs), R1 (7 yrs) & B2C & $>$ 100,000  \\ 
O7 & U2 (3 yrs) & B2B & $<$ 100\\ 
O8 & U6 (2 yrs), D2 (4 yrs) & B2B & $>$ 100,000 \\ 
O9 & S3 (1 yr) & B2B & $<$ 100 \\
O10 & D3 (6 yrs) & B2C &  1,000 -- 5,000 \\ 
O11 & U3 (1 yr) & B2C & 10,000 -- 50,000\\ 
O12 & U4 (1 yr) & B2B & 100 -- 500 \\ 
O13 & S6 (2.5 yrs) & B2B & 5,000 -- 10,000\\ 
O14 & S7 (3.5 yrs ) & B2C & $<$ 100\\ 
\end{tabular*}
\vspace{5px}
\caption{Each organization is listed with interviewees by role (S = Software Engineer (AI), U = UX Professional, M = Manager, D = Data Scientist, R = Research Scientist) and a brief description. The number in brackets next to each interviewee indicates the participant's years of professional experience in HAI.}
\label{table:organization_interviewees}
\end{table*}


\subsubsection{\revisions{Procedure:}} \revisions{We conducted interviews with 21 industry professionals from 14 different organizations of differing sizes (see Table \ref{table:organization_interviewees}). Each participant was interviewed separately (i.e., we conducted 21 interviews in total). We recruited individuals involved in building AI components for user-facing products; mainly, UX professionals and AI engineers, data and research scientists, and managers. Starting with university alumni and other industry connections, we used snowball sampling to recruit participants through referrals. They had between one to 12 years of professional HAI experience, with 13 having at least three years and an average of 3.7 years (SD=2.5 years) (Table \ref{table:organization_interviewees}).
Participants were not compensated for participation, but could opt-in to receive a small gift of university merchandise. Before the interview, participants completed a consent form, and in many cases, also sought approval from their company's legal team. The first and second authors conducted all interviews through video-conferencing, with each interview lasting about 60-minutes.}

\revisions{In these semi-structured interviews, we started by asking about the participant's role within their company and their team affiliation and organizational structure. We then asked them to choose and describe an AI-based application they helped create. We asked participants to walk us through how they participated in the process of creating the application (as allowed by disclosure rules). Based on their description, we used follow-up questions based on our component model about whether (and how) they operationalized different guidelines, different roles involved in the process, and their workflows for collaborating with others. For example, we asked designers how they learned about potential AI errors and asked engineers how they obtained requirements for the particular feature they built. We also inquired about conflicts during the collaboration and how they were resolved. Participants were probed about the kinds of tools they used (e.g., in-house tools versus outsourced ones), whether and how they referenced existing Human-AI guidelines, and if there are any tools they wished they had. Later in the interview, we inquired about their use of prototypes. Other questions addressed perceived differences and similarities between AI-driven and non-AI applications. Finally, we asked participants for their feedback about Human-AI design guidelines and ideal workflows for collaboratively building AI-based applications. The interview questions are available in the supplementary materials.}

\subsubsection{Analysis:}
We contracted a third-party service to transcribe the interview with the exception of one where the team manually transcribed at the participant's request. We then conducted qualitative coding analyses  using a grounded theory approach \cite{strauss1990basics} starting with an initial review of the interview notes and an \textit{in-vivo} analysis. Two authors independently open-coded five transcripts and then collaborated to develop an initial codebook, resolving disagreements by consensus. The resulting codebook consists of 40 top level codes including the use of prototypes and artifacts, multiple workflows, friction or tension in collaboration, differences between AI-driven apps and conventional software, and tools used for communication and collaboration. The complete set of codes is available in the supplementary materials. The two authors then individually analyzed the remaining transcripts using this codebook~\cite{denzin2011sage}. Because we used a grounded theory approach, we did not see a strong need to compute coder reliability \cite{mcdonald2019irr}. A memoing activity synthesized findings across transcripts~\cite{birks2008memoing} focusing on how collaborative teams develop human-AI applications.

\subsection{\revisions{Findings}}
\begin{figure*}[t!]
\centering
\includegraphics[width=\textwidth]{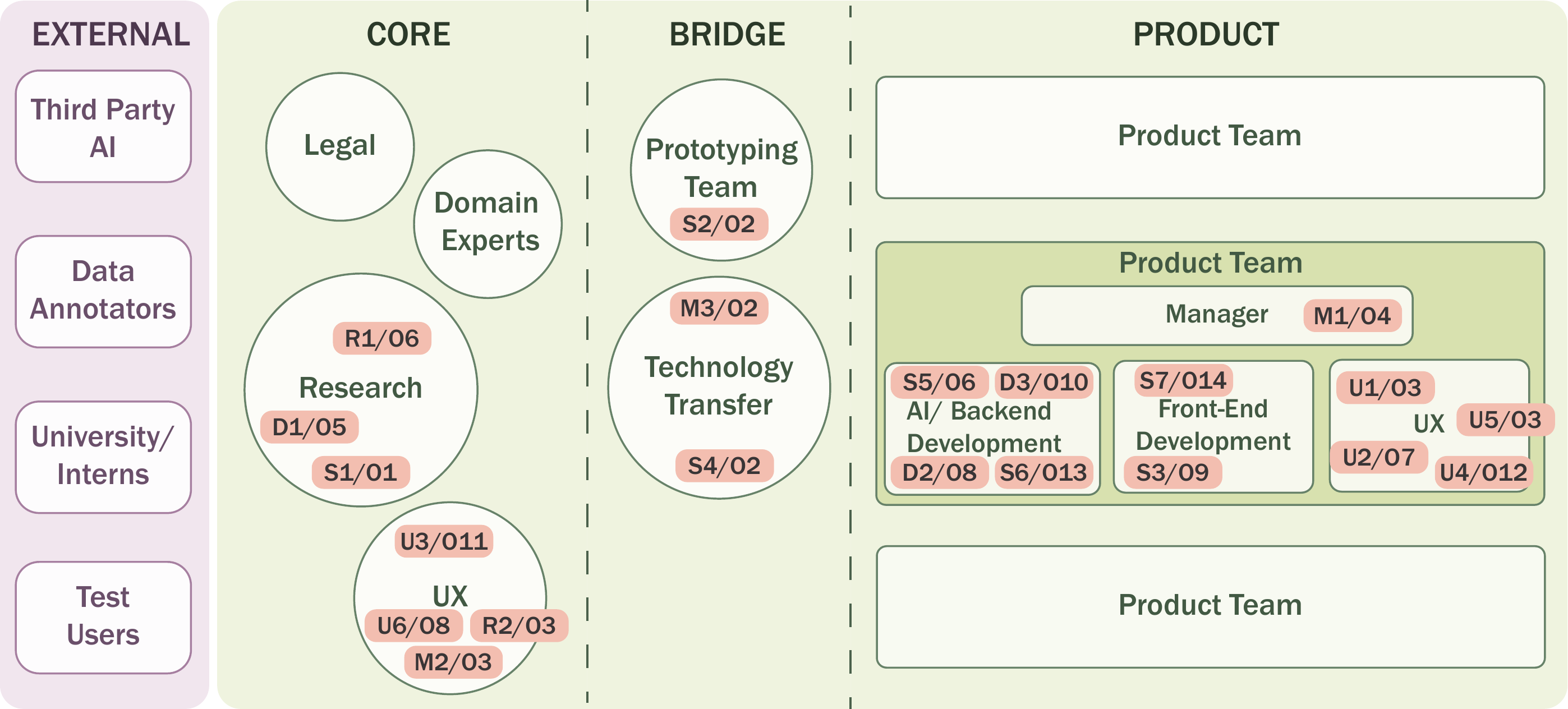}
\caption{Generalized Organizational Structure of Teams in Human-AI Application Design and Development. Interview participants are overlaid onto corresponding teams (S = Software Engineer, U = UX Professional, M = Manager, D = Data Scientist, R = Research Scientist). "O" denotes the organization number.}
\Description{Visual representation of organization structure for HAI software development. The high-level structure includes core teams, bridge teams, product teams, and external entities.}
\label{fig:structure}
\end{figure*}

Our interviews reveal how team structures and separation of concerns (boundaries) between differing roles and expertise \textit{hinder} human-centered AI design. Several of our participants reported a separation between individuals who \textit{conceptualize} AI capabilities and those who \text{integrate} those AI capabilities within end-user products. As shown in Figure~\ref{fig:structure}, many large organizations have dedicated AI research teams (primarily computer scientists) who explore novel AI capabilities and techniques. In these teams, the focus is on advancing foundational research in AI. Product teams are not typically involved in this process, and the technology itself may be only partially motivated by real end-user needs. However, once the technology vision is achieved, research teams join in with different product teams to identify product use cases for applying AI innovations (i.e., an AI-first workflow). 

To support the research-to-product pipeline, as reported by three participants, large organizations may have intermediary technology transfer teams that envision product and human uses for AI innovations. On the other hand, smaller organizations and start-ups may rely on third-party AI providers (e.g., Microsoft Azure AI~\cite{azure}) to add new AI capabilities into product features. Outside of core research and product teams, AI development commonly requires support from domain experts and data annotators. These teams tend to be external to the organization. Further, according to two participants, teams consult with legal representatives about ethical data collection and data privacy issues. Both large and small organizations may have a pool of beta-testers available to evaluate new features during development. Collectively these team boundaries introduce numerous challenges to operationalizing the HAI design guidelines. We summarize our findings in terms of (1) limitations due to separation of concerns at the design-engineering boundaries, (2) design workflow challenges from data centric nature of AI development, and (3) current workarounds to alleviate collaboration difficulties at the boundaries.

\subsubsection{Design-Engineering Boundaries Hinder the Cross-Cutting Requirements of HAI Guidelines}\hfill

\textbf{Boundaries Introduce Knowledge Blindness about End-Users and AI Capabilities.} 
HAI guidelines recommend that the AI capabilities should be motivated by human needs and should align with human behavior, task workflows, and cognitive processes (see Figure~\ref{fig:model}). However, the boundaries between core AI developers and UX designers limit possibilities for creating human-centered AI from the \textit{ground up}. Given the novelty of AI, researchers and engineers are motivated (and incentivized) to explore AI capabilities independently and without regard to products and end-user needs. As manager M3 described: \textit{``\ldots research coming up with new cutting edge state-of-the-art techniques for doing something that the product team wasn't even thinking about, or users aren't asking for, because they hadn't thought that way.''} This boundary separates core AI developers from end-user product teams and introduces \textit{end-user blindness} about product users' needs and concerns. The result is often erroneous assumptions about what users would want from AI. In describing their frustration due to end-user blindness, manager M1 commented: 

\begin{quote}
   \textit{ ``You have these requirements where you need these videos to be analyzed, and tagged, and categorized\ldots a lot of times people [AI engineers] would go off in just weird directions, get obsessed with trying to identify guns, or something that wasn't really that important to what we were doing''} - [M1]
\end{quote}

On the other hand, product teams---specifically UX designers who advocate for end-users in design specifications---may lack an understanding of AI technology capabilities, i.e., \textit{AI technology blindness}. As a result, UX designers appear to either distrust or over-rely on AI capabilities,  which manifests in their UX design for AI. As research scientist R2 puts it, designers tend not to automate things that they could be automating: \textit{``There's under trusting where it's like oh actually you should let the algorithm make a suggestion, maybe offer a choice, maybe you should trust it more than you do.''} R2 further adds that in other cases, there is over trust on what AI can and cannot do: \textit{``\ldots then other times, especially when you get into the cases around anthropomorphism and things like that, people really overshoot their trust and think yeah this is going to be great no matter what happens\ldots''} A consequence of the black-box approach to design is that designers themselves lack clarity about the AI performance and output. This makes it challenging to design user experiences that align with end-user mental models and expectations. In advocating for overcoming technology blindness in UX design, M2 commented on designers needing to understand the capabilities and limitations of AI: 

\begin{quote}
    \textit{``It used to be that UX designers made static mocks and there was screen-to-screen flow. But with AI the probability that the end user makes it through that path is lower because it is driven by a sequence of machine learning models. Just the probability curve on any journey is now much more complicated and is much more branched. Designers need to understand probabilities. They need to understand the failure cases, understand confidence and how to deal with confidence scores, how to interpret thresholds. They need to be able to understand the grain of the technology\ldots the possibilities and the edges.''} - [M2]
\end{quote}

\textbf{Traditional Software Practices at Boundaries Impose Premature Specification of AI and UX Designs.}
In conventional software design, UX professionals work with end-users to define the requirements for the application. However, because of knowledge blindness, on their own, designers and engineers may not be able to fully define specifications for either the AI model or the user experience and interface. Yet, our interviews identified the tendency to define AI and UX specifications \textit{independently} because of the work-role boundaries and lack of coordination practices. This problem takes control away from designers attempting to craft the end-user's experience. Across many interviews, designers expressed frustration in trying to design the UX around an independently created AI specification. For instance, in one of the sessions, the AI team developed a nuanced tagging schema for media content and handed it off to the UX designer to integrate into a voice assistant. The designer (U1) commented on the challenge in integrating UX around predetermined AI specifications, noting the extensive rework and feedback required to retrain the AI tagging model in a way that meets their understanding of end-user needs: 

\begin{quote}
    \textit{``In the first meeting, [the AI team] were just saying `We need to add this [categorization] on the screen, can you find where is the right place?' Then I need to work a little bit backward to say there are constraints on wordings, and voice UI had another layer of constraints, so I need to understand how this information fits into the screen we already have. They have their own taxonomy on what kind of information they are looking for, but for users, it doesn't evoke a response~\ldots The [label] semantics is not on the same level to what we already have.''} - [U1]
\end{quote}

Similarly, AI engineers also found it challenging to implement desired AI features when the UX is already specified in great detail without AI involvement. AI models (unlike conventional applications) are challenging to build to specifications because their behavior is dynamic. This makes it difficult for engineers to independently create AI technical specifications from design requirements alone. S7, an AI engineer, commented about their frustration in the coordination and hand-off process between UX design and engineering: 

\begin{quote}
  \textit{ ``\ldots they [UX] would hand that [design document] off to the engineer and say `Implement this.' And of course my reaction to this was `This is garbage.' This does not reflect the appropriate architecture for implementing this thing. It felt particularly extraneous when it got very granular, and it was not the best medium for describing the desired behavior. Because the designers were not technical really. This is not a good reflection of how the actual AI software engineering is going to happen. And I was like, `Stop trying to do my job for me.''' } - [S7]
\end{quote}

The problem of AI blindness among designers arises from the role boundary created by professional expertise. By advancing UX design independently from AI teams, UX features become ``set adrift'' from the other source of constraints for end user's needs---the AI model.

\textbf{Boundaries Limit Access for AI and UX Collaboration.}
Because of differences in the design and engineering processes, there is no clear understanding of how human-centered design requires alignment across both tasks. For instance, U6 expressed concerns that the UX team was not involved in the training data annotation process---the core of how end-users experience the AI. According to U6: \textit{``it seemed very odd to me that as designers we were not invited to the annotation session. So, we had to invite ourselves to just talk to domain experts\ldots''} U6 further commented that \textit{``\ldots for engineers, their approach is more like `the machine is going to figure it out.' We could be talking about health or elephants in the circus, and it is all the same to them \ldots'' } 

Across the interviews, other collaboration challenges also emerged. First, the core responsibilities for UX professionals are defined differently from basic AI research. In addition, the time needed to conduct user research was viewed as out of sync with AI research progress. For instance, U4 comments that \textit{``we don't necessarily participate as much in that whole AI thing, but the thing is because we're also trying to make sure that we're doing user research and participating in that.''} S4, a research engineer in a different organization, offers their perspective on collaboration:  \textit{``\ldots they [UX] might complain after the fact that they weren't early enough, but on the flip side if we try to involve early then they'll say they're busy doing x, y, and z. In my experience, it's not always practical.''} Acknowledging the added time it takes to conduct user research, M3  comments: 

\begin{quote}
   \textit{ ``You obviously need a human need, or you're not going to have anything worthwhile, but the reality is in most of these companies there are in-flight research efforts that are happening on basic capabilities, and it's not like you can say, `Okay, everybody stop what you're doing until I find you a human need, and then you may start working again.' It's just kind of absurdity.''} -[M3]
\end{quote}

Further, in smaller organizations that work with third-party AI services, boundaries severely challenge the design of AI behavior and presentation for end-users. In working with third-party AI, M1 describes that designers often have to engage in the laborious activity of translating AI output into user-friendly labels and formats: \textit{``\ldots the label that the database has for the data may not be the same as what your end-user understands it to be. So, understanding there's a difference between how an engineer labeled it in the database, versus how you might want to show it on your UI to the end user\ldots we would look at the raw JSON files and create our own labels \ldots''} The disconnect between  external AI services and products forced designers to alter the AI model specifications to avoid AI experience issues for end-users.
\subsubsection{AI's Data Needs Transform Designers' Role in HAI Development} \hfill

In conventional UX workflows, designers synthesize higher-order system requirements from end-user data. However, in HAI workflows, individual data points (i.e., examples) \textit{are} essential requirements for AI development. Our interviews revealed constraints to human-centered data requirements and designers' involvement in AI development pipelines. 

\textbf{From Developer-Centered to User-Centered Data Requirements.}
In contradiction to guideline recommendations, in AI-first workflows, technology requirements appear to drive data requirements. When exploring new AI capabilities, researchers don't always know what types of data might be needed. Requirements about data and its characteristics, such as variables, data types, labels, and the number of data points, evolve through a ``trial and error'' approach. As reported by three participants, this workflow aims to optimize the AI development process. That is, researchers start with an initial, small dataset to train the model. For early-stage data collection, organizations may have internal data collection and logging apps (e.g., one that collects gesture data while using the phone) that can be deployed across different teams. There is often an ``unwritten agreement [S5]'' that development teams will provide data for AI development purposes. This lowers the cost of data access. As S5 comments: \textit{``\ldots you have to spin up a dedicated user study program, go through a lot of process, a lot of review, it's a whole lot of bureaucracy to get that kind of rich data collection.''} Therefore, AI researchers work with pre-existing data sets or  `minimum-viable-data' collected from within their UX team and then gradually increase scope of data over time:
\begin{quote}
   \textit{ ``For collecting data, we will start from people within our team and do a first pilot testing [of the AI]. If it works well, we will increase the size of the data. For example, we will recruit people from a different project team so they are not biased to our data needs. And if it continues to work well but we still need more data, we will start looking for external partnerships to collect data from our target users.''} - [S1]
\end{quote}

In this workflow, engineers also reported striving for ``clean'' data by removing outliers and noise to improve model performance. This may lead to an idealistic version of data for AI model exploration that omits data characteristics and features relevant to real-world use.

Once the AI capability and data specifications are determined, UX researchers get involved in validating data needs with end-users or collect additional data to test for AI robustness. For instance, UX teams may work with end-users and customers to validate data labels for the application design. As M2 comments: \textit{``if you want a certain data structure with a hundred hypothetical labels, you can show that to users and get sentiment on that\ldots''} Further, UX designers commented that such a partnership requires careful consideration about privacy and content ownership, as well as communication about benefits to customers. AI engineers also emphasized the need for clear communication about how customers (who are assisting with data labels) might benefit from their contributed data. Because of the way user inputs are elicited, S6 commented on end-users being hesitant to provide information for labeling tasks:

\begin{quote}
\textit{ ``We asked customers [to label the data], but it wasn't good enough for our use. Anecdotally, I think the people who are being asked to label weren't sure how this information is going to be used. I think there was some hesitation because it wasn't tied to their day to day metrics or goals. I don't know if there was an element of fear of automation\ldots''} - [S6]
\end{quote}

As a consequence of AI model needs, end-user data collection appears to occur more incrementally and less formally than in conventional applications. Further, this alters how designers conduct user research to now include the AI development pipeline. 

\textbf{Designing Data Collection Tools with People in Mind.}
Given the significance (and multiple roles) of data in HAI design, data collection and annotation tools are essential for gathering end-user requirements. Consequently, engineers develop custom tools for collecting needed data. According to S1: \textit{``A lot of times, our problem is not generalizable, so we build our own tools in-house.''} Such tools are often optimized to lower the engineering cost for data cleaning and transformation. For instance, the data collection tool may explicitly ask participants to start a session and perform some task or prompt participants to validate whether or not the correct label was detected (e.g., labeling sensor-based activity detection). Both designers and engineers acknowledged that labeling could be tedious work. They expressed empathy for people charged with labeling the data (e.g., \textit{``there are overseas sweatshops where people are just filling in Mechanical Turk surveys day in and day out, figuring out whether the image has a dog\dots with all the empathy in the world you have, you feel really bad for those people''} [S2]). In one interview, the designer reported visiting those performing labeling on-site to understand their pain points and run user studies with them to evaluate data annotation tools.

\begin{quote}
    \textit{ ``We wanted annotators to create object segmentation boundaries on images by drawing polygons. To design the tool, I visited [location] and asked the annotators to generate labels. From these trial runs, we noticed that using the keyboard was essential for a good UX, and they needed ways to undo or edit polygons. Based on this, we did a focus group to know how we can improve the labeling tool.''} - [S3]
\end{quote}

This example illustrates the change in the nature of user research data,  how it is collected to design HAI systems, and how it is used. Designers are learning to provide new types of UX support driven by AI model development. Their objective is to optimize the user experience for end-users but also lower engineering effort in data usage.  

\textbf{Authentic Data for HAI Evaluation.}
As with the technical evaluation of AI models using a ``holdout'' dataset, in many instances, human-centered evaluation requires that designers adapt evaluation practices for end-users to supply their own data based on their personal experience history in a domain. This allows end-users to provide feedback about AI behavior from their viewpoint as experienced within their own situated contexts. As R2 puts it: \textit{``The best mock for AI is a lot of times human. We really try to use people's own content. This is the thing; if I look at photos of my friends and family, I'm going to have an emotional reaction, I'm going to have an authentic experience there.''} Consequently, AI model design requires continued evaluation and feedback from diverse end-users with personal experiences in a task domain. However, constant engagement with end-users (ranging from novice to domain expert) within existing design and development workflows is challenging for UX designers to accommodate. In describing this challenge, S5 comments:\textit{``User studies, especially things of this nature, like, getting around a lot of our privacy constraints tend to be difficult, which that's a whole another like, can of worms you probably don't need to attack right now.''} S5 points out that evaluation, especially for recommendation systems, requires access to user data and requires time-consuming review for privacy compliance.  

In addition, teams find it challenging to develop the right metrics to gather feedback on AI experience design. According to D3 \textit{``To me, evaluation is still very, very hard. And especially I think maybe more subjective evaluation too in terms of the quality or how enjoyable was the experience?\ldots if you were using the measure of how many items you interacted with or how long you engaged, it would feel like the one that was a five-item engagement was more successful than the two-item engagement, where actually they [end-user] didn't really think that at all.''} (D3). A lack of well-tested metrics makes it hard to run deployment studies to gauge end-user expectations and trust. These challenges are amplified in evaluating AI behavior over time, especially for learnability through end-user feedback. Addressing these issues will require designers and engineers working together to identify appropriate performance metrics and privacy-preserving evaluation strategies. 
\subsubsection{Bridging Boundaries Through Collaborative Design and Constant Co-Evaluation}\hfill

In responding to the expertise boundary and data role challenges, participants revealed how they reduced friction to facilitate engagement across teams. These workarounds involved a variety of boundary negotiation artifacts~\cite{lee2007boundary} for (1) knowledge sharing, (2) collaborative prototyping and design negotiation, and (3) design evaluation and feedback.

\textbf{Bridging Knowledge Boundaries between Designers and Engineers.}
In conventional software workflows, UX designers rarely share raw end-user data and low-fidelity representations with engineers. However, the interviews revealed that sharing low-fidelity artifacts is effective in centering the end-user within AI model design. For instance, UX designers reported sharing raw user data and co-creating user personas with engineers to help them think about training data needs. While it demands a more extensive data collection program, collaborative synthesis generates requirements for different data collection tools, types of end-users to recruit, storing and processing data, and collecting data preserving privacy and ethical concerns. In describing their approach to ensuring the representativeness of different end-user groups in collected data, U5 comments: 

\begin{quote}
    \textit{``Often, I look out into the world to see what information is there about existing groups, and then evaluate for myself, do these groups make sense or do I need to make new groups. I have done all of the user research and come up with groupings on my own and then brought them back to the team. Then I talk it through with the PM and the engineers what the value of different user segments are, why would we want to prioritize the different users, why are they important to the company\ldots’'} - [U5]
\end{quote}

With HAI, the task of anticipating relevant differences in end-user populations impacts not only the UX design but also the behavior of the resulting AI model through training. Another change in UX designers' work for HAI occurs when designing interaction or task workflows. AI engineer S7 reported that designers sharing sketches and storyboards (instead of high-fidelity prototypes) offered flexibility and control in mapping user needs to AI features and implementation logic. 

\begin{quote}
    \textit{ ``Storyboards or other documents that get into describing what the purpose of the behavior is, what the desired user experience is without getting into the engineering. I think of it as a sort of comic book illustration of what the user experience should be and what the system's reaction should be in different interactive situations. It was like sort of the key expected traversal through an interaction, and then maybe some of the most likely other paths about what experience you want the user, and the [system] to have together. Here is a situation, and what should happen over the course of this interaction. And I don't mean to seem territorial about this, but it's really useful to have back and forth with the people who are trained to think carefully about user experience\ldots''} - [S7]
\end{quote}

Similarly, engineers reported varied strategies for sharing AI capabilities and details about implementation (such as assumptions and logic) with UX designers and domain experts on projects. Again, the intent is to resolve technology blindness and to facilitate collaborative design and feedback. As S6 comments: \textit{``If we don't adequately communicate to designers, they fill in the gaps with their own theories and its not clear what input needs to be provided in order to get the desired results.''} In one scenario, AI researchers reported working with university interns to develop a conceptual prototype of an AI feature. Here the goal was to (1) demonstrate a new capability of AI within an application context and (2) define a design space for UX researchers to think about the experience. As S5 describes: \textit{``We got something tangible enough that we could actually go talk to a designer and\ldots we started letting them play around with it, and said, `Try it out for a week and tell us is this better than the old way that we've done things.'\ldots it also broke the problem down such that the designers understand, here's the benefits of where the machine learning can be applied.''} Once UX designers understand the AI design space, they are able to collaborate with researchers to explore end-user needs, using the prototype as a design probe.

In other cases, AI researchers may identify a new technical capability but find it hard to define its use context. In such cases, UX researchers need first to understand the technology and then identify its benefits for potential end-user experiences using prototyping approaches (as suggested earlier). As M2 described: 

\begin{quote}
    \textit{``A lot of times, people are, just kind of, down in the weeds, really deep and get a little lost in the day-to-day work. UX teams can actually bring a little hope to those folks and give people a target, and really paint a picture of that through design visualization, whether that's making a movie or just making a series of mocks, or building an experiential prototype, or something like that, really help land the tangibility of something that's pretty deep and complex. Sometimes, it's the light at the end of the tunnel\ldots''} - [M2] 
\end{quote}

In this regard, two participants, both project managers, emphasized the value of UX-friendly machine learning tools for creating experience prototypes. Specifically, these tools allow designers to take ``off the shelf'' ML models or plug in their own team's model and work with real end-user data to demonstrate an envisioned AI feature. In addition, using functional ML models mitigates the danger of setting or communicating unrealistic expectations with AI mock-ups. 

\textbf{Collaborative Design of HAI Prototypes.}
By bridging expertise boundaries, designers and engineers reported working towards collaborative prototyping, including data and labels, AI model behavior, implementation, and end-user experiences. For instance, HAI guidelines recommend defining data labels and annotations by consulting with expert users. In the interviews, UX designers and engineers identified multiple ways to work with domain experts to co-design labels. For example, in one interview, data scientist D3 reported that they generated database queries for exposing different views of training data requiring labeling. Engineers then define ML constraints for labels, and UX designers and domain experts generate and validate labeling schemas (i.e., rules for assigning labels to raw data). A second example occurred when data scientists find pre-existing datasets they need to re-purpose for their AI needs. In this workflow, data scientists work with domain experts to clean data, identify variables for prediction,  interpret data analysis results, and perform labeling. As D2 describes: \textit{``we would be talking to meteorologists about how to adjust variables, and create flag variables, so if it is above this temperature or dew point, we would categorize it\ldots''} This collaborative process happens through sharing CSV files, Python scripts, and visualizations.

Further, creating experience prototypes combining AI capabilities and UX needs requires close collaboration between designers and engineers. In the case of Wizard-of-Oz prototypes, UX designers gather end-user data and work with engineers to generate outputs and understand the logic behind them. This is essential to understand the unanswered questions from an engineering standpoint, plan the type of user study needed, and design the presented experience of the prototype for end-users. As U3 describes: 
\begin{quote}
    \textit{``Let us say I am doing food recommendations. And I want to tell users why something was recommended. It may be because they are liking a few restaurants, or they added items to their shopping cart, or maybe it is because of past orders. It is a Wizard-of-Oz prototype where I first get users’ data. Then I get the model output from the data scientist and work with them to understand the model labels and explanations. The data scientist wrote down all the equations and explained it to me very clearly. They showed me how the weights were set, and we discussed things we need to know from users, whether to do an A/B testing or a walkthrough\ldots''} - [U3]
\end{quote}

Engineers also support UI designers through annotations on UI wireframes about what is happening behind the scene. According to S6: \textit{``I added annotations on the side about what is happening behind the scenes like an API is being called. Then as an example, I would [annotate] for the API what output it comes back with\ldots I use Balsamiq [UI prototyping tool~\cite{balsamiq}] because I think it lowers the barrier of what can be a design tool, and you don't need specialized knowledge to communicate that idea.''} These comments by developers indicate efforts to support greater collaboration and extension of expertise across boundaries.

\textbf{Design Iteration with Constant Evaluation.} \label{sec:design-iteration}
Lastly, the interviews revealed that evaluation happens frequently using unfinished prototypes still under development. Participants reported that this form of assessment is necessary when the user experience design is co-evolving with AI development: \textit{``I think the process that works best is fairly tight review cycles with the actual evolving behavioral artifact.''} (S7).

\textit{Early stage evaluation of model behavior.} In the early stages of development, engineers may make certain assumptions about AI behavior. Frequent evaluation allows UX researchers to provide early feedback about these assumptions. As S7 describes: \textit{``\ldots as I was implementing this feature and I ran into this problem of how to handle this use case? \ldots Here is the guess that I made, but let us talk about whether that was the right choice. As things were getting built, we would look at the running prototypes and be like, `Do we like how this plays?'\ldots''} Here, S7 describes how this approach is more suitable for AI development compared to having a black box prototype provided by the UX designer. Similarly, for AI perceptual interactions (e.g., computer vision), UX designers may supply an initial set of desired interaction gestures. Then, during development, designers and engineers evaluate the feasibility of those interaction gestures using prototypes and discuss alternatives. S1 explains that:

\begin{quote}
    \textit{``The designer will say `we want ten different facial expressions for this model'\ldots we start from there to build the backbone of the interaction, and then we iterate through it\dots we call like grayboxing\ldots there are three facial expressions where it's just really hard to get that right, it is not going to perform very well. The other seven is fine. So, in the process of testing, we find out, there are two other facial expressions that are not in the original ten expressions that can perform pretty well. And so we will tell the designer that these three we will need to cut it. But if you want, there are two more gestures, you can add into your interaction\ldots''} - [S1]
\end{quote}

In a different scenario, evaluating early prototypes with target users helped engineers determine the optimal algorithm for a problem (user need) they are trying to address. In describing the iterative process of model comparison to find the best approach, S2 explains that: \textit{``\ldots the process involved 20 different prototypes I had to build for all the different algorithms we've tested on.''} Further, they describe that the prototypes expose the actual model logic using visualizations for test users to evaluate:
\begin{quote}
    \textit{``\ldots user uploads an image, and I visualized the palette for the image so that we know how the algorithm is working under the hood. Because AI is a black box, we need to have some transparency for the user to understand. You show the palette. Once you have the palette, it will search and return the results. For each result, I also show the pre-indexed pallets we use to compare with other algorithms.''} - [S2]
\end{quote}

This allowed the UX designer to do a comparative evaluation \textit{iteratively}: \textit{``Every week when we have a new algorithm, we compare it to the existing best and see which one is still the winner and then that will compete with the next algorithm. \ldots That is how we reach to find the one which we shipped to productions.''}

\textit{Iteration with domain experts to determine AI behavior and interpretability.} In other cases, the data scientist may provide domain experts with a spreadsheet containing rules and assumptions made in building the AI model. The expert then annotates changes to the rules for updates of the model. According to D1: \textit{``There are rules and codes we have that we use for making recommendations. We would list out the rules so the domain experts could look at them. Then, we started to give them more accessible tools like sharing a spreadsheet where they could give their inputs\ldots They could flag, add notes and annotations [about model output].'' } This process allows domain experts to participate in specifying AI behavior at a conceptual level. Similarly, designers worked with engineers and domain experts to evaluate interpretability features by creating functional prototypes with different output formats. For example, U2 discussed their process for showing output probability to end-users, working with domain experts to translate percentages into categorical bins, such as ``high'', ``medium'', and ``low''.  
\begin{quote}
    \textit{``There are two versions we iterated. The first one is to show the possibility as numbers. If I have ten patients and nine of them have 100 percent, and only one shows 20 percent, it might confuse a user because a number is really hard for [end-users] to understand\ldots The second version we actually tried was high, medium, low possibility. So that turns out to be more positive by the user.''} - [U2]
\end{quote}

\textit{Evaluating data and model for privacy and ethical concerns.} To evaluate privacy and ethics concerns during data collection, AI engineers often collaborate with legal team members. Many interviewees described this as a collaborative process where engineers walk through what data is collected and why. Then, they discuss alternate data sources in case of privacy violations and how to collect data in a privacy-preserving way. As described by S5: \textit{``all data collection has to go through a privacy review \ldots you sit down with one of them, you walk them through, here is the data we want to collect, here is why we want to collect it. Then, they discuss about is all this data necessary, can we do different ways to interpret it?''} This process often involves sharing compliance documents and details about model implementation and data, and a legal team may draft a privacy statement for end-users to review.

\textit{Evaluation in the wild.} When a fully functional prototype is available, UX researchers may conduct deployment studies with test users to evaluate how the model performs in the real world. M2 describes this process as \textit{``Anybody can download [the] app and try it out, That's how we collect data a lot\ldots it's very easy for a UX researcher to go back and say, `We see this fail for this use case,' or `for this population,' and just go back to the team and it's an open conversation about the limitations of the current model and how to adapt\ldots.''} Further, UX researchers may conduct a longitudinal evaluation with functional prototypes. According to U5: \textit{``\ldots doing longitudinal research is really helpful \ldots if it is something that takes a bit of ramp-up time, giving the people you are testing with time to spend with it, to see where it lands and how useful it is over that time.''} In communicating to users about longitudinal testing, U5 comments that:

\begin{quote}
   \textit{``I think some of it is just product transparency, it would make sense for me to just be like, `Right now we don't know anything about you, but come back as you use this app over the next couple of weeks. We will start to produce better recommendations for you.' So keep checking back because otherwise, I think you might make assumptions that it will never work or things like that. So I think transparency can be really helpful in those situations.''} - [U5]
\end{quote}

One challenge with this iterative process is communicating with designers and end-users about what is implemented (and what is not) and what type of feedback they need to provide. Identifying primary functions to test, and why, along with which functions are missing and why they don't matter at this moment, requires UX designers to have a high-level understanding of the AI development process. As S7 puts it:

\begin{quote}
    \textit{``I think that part of being in a nontechnical role is understanding enough about development. So you need to tell them `Listen, what we are showing you today is two weeks of work. Here are the things that it doesn't have but it will have. We don't need feedback on the fact that it doesn't have sound effects or graphics. What we need feedback on is, is this the basic kind of interaction you want? Does this look like something that is going to solve the problem? Trust us. We will get back to polishing it, that is not what we are looking at at this stage\ldots.''} - [S7] 
\end{quote}

In summary, we find that teams overcome collaboration challenges in HAI design by disregarding conventional software separation-of-concerns to create and share low-level design and implementation details across knowledge boundaries.

\section{Discussion}
\revisions{Our findings show that SoC introduces numerous challenges to HAI software development. First, we find that \textit{delayed specifications} in software workflows is integral for operationalizing the HAI guidelines. As Yang et al.~\cite{yang2020re,yang2019sketching} note, UX professionals lack familiarity with AI capabilities and the means to design AI components. This leaves key AI specifications (such as feature selection and model assumptions) to those with technical expertise (echoing Zhang et al.~\cite{zhang2020data}). Our study especially builds on prior work by finding that AI specifications are often made prematurely, necessitating difficult and expensive changes when user experience concerns are later identified. Changing workflows to \textit{postpone} technical commitments may facilitate the continued integration of concerns throughout the design process. Next, we report task-specific creative workarounds introduced by both designers and engineers to overcome knowledge blindness and support collaborative HAI design. Importantly, we concretely show \textit{how} HAI teams can achieve multidisciplinary collaboration. These workarounds contradict established software development practices dictating abstraction, information hiding, and modular design in SoC. Here, we further characterize these workarounds as \textit{leaky abstractions} intended to share key information across concerns. Leaky abstractions appear to help designers and engineers (1) coordinate specification and implementation details, (2) collaborate on designing both the AI and UX, and (3) integrate concerns across disciplinary boundaries to develop HAI systems. Building on leaky abstractions, we theorize towards a collaborative design workflow through delayed specifications and constant evaluation. }

\subsection{Leaky Abstractions \revisions{Support} Collaborative HAI Design}
\revisions{In collaborative software design, the purpose of abstractions is ``not to be vague,'' but instead to ``create a new semantic level in which one can be absolutely precise''~\cite{dijkstra1972humble}. However as our findings show, existing abstractions that are effective for conventional software development, hinder collaboration in HAI system design. Moreover, given the recency of HAI guidelines and its novelty for software practitioners~\cite{yang2020re,cai2019software}, our understanding of abstractions for design-engineering collaboration is still evolving. Our definition of ``leaky abstraction'' as ---ad hoc representations shared across expertise boundaries to expose low-level design and implementation details---captures this evolving nature of understanding over time. Through a related characterization of representations as technical objects and epistemic objects \cite{ewenstein2009knowledge}, leaky abstractions emphasize the importance of incomplete and constantly changing design knowledge during HAI software development. Leaky abstractions also encompass a wide variety of inter-designer representations~\cite{visser2006cognitive} for collaborative design. We observed instances where teams created different leaky abstractions to address explainability, error handling, feedback, and learnability. As Yang's description of ``designerly abstractions and exemplars~\cite{yang2018investigating}'' suggests, leaky abstractions may be needed to address a wide variety of cases where lower level detail is needed to inform and support interface and interaction components. In our interviews, leaky abstractions took many different forms and addressed many different elements of user interface and AI model design. }

\begin{figure*}[t!]
\centering
\includegraphics[width=\textwidth]{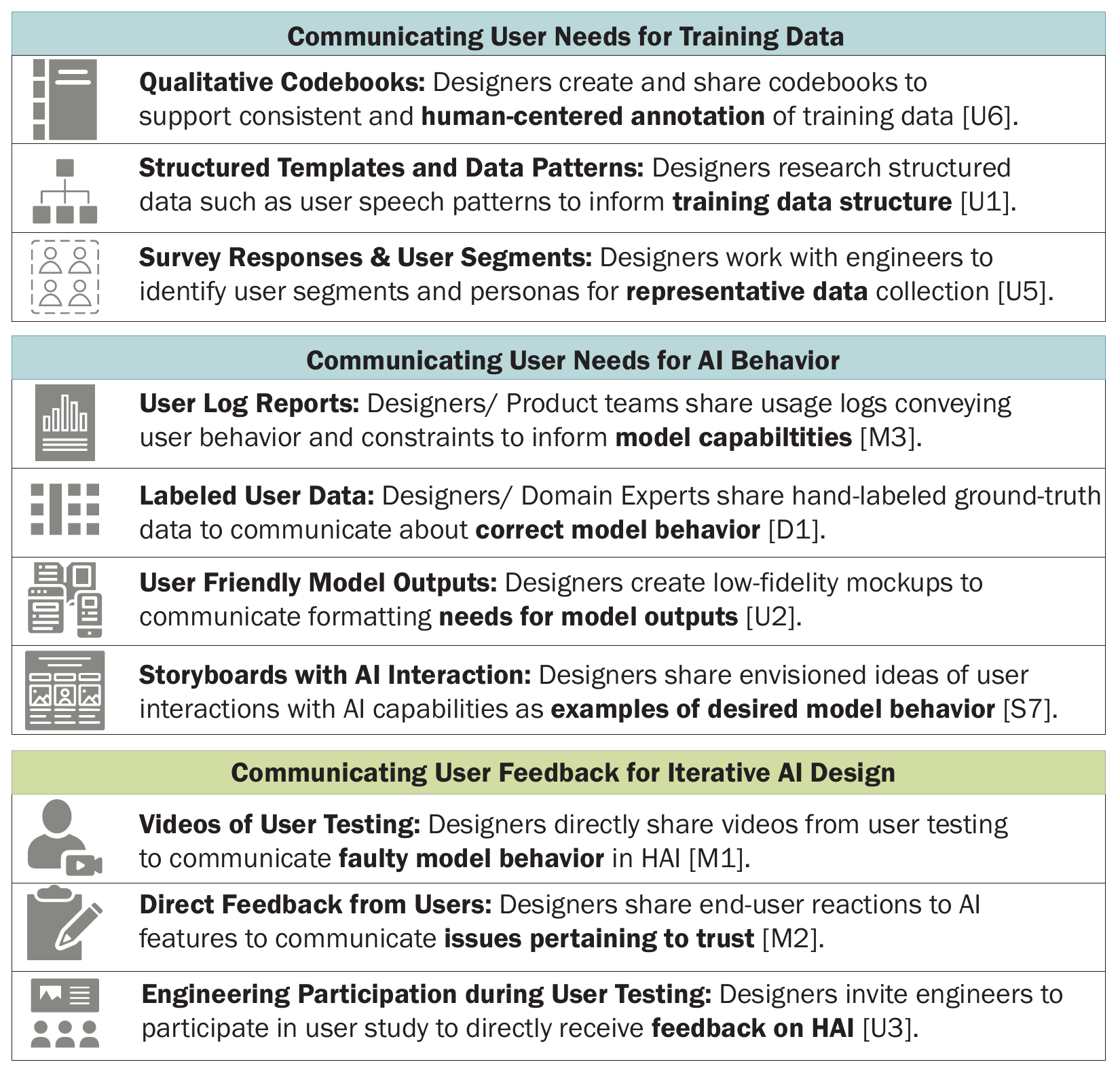}
\caption{Designers Share Low-Level UX Knowledge with Engineers to Inform AI Implementation Decisions}
\Description{Table of leaky artifacts shared by UX designers with AI engineers for communicating training data needs, AI behavior, and user feedback for iterative design.}
\label{fig:deleaks}
\end{figure*}

As reported in our findings, designers shared low-level design details with AI engineers to shape AI around end-user needs  (see Figure~\ref{fig:deleaks}). First, contrary to conventional wisdom, designers shared details about personas and user segments that emerged from surveys, qualitative code-books for training-data labeling, and raw end-user data (gathered through UX research processes) to inform representativeness and formatting needs for AI's training data. Second, designers shared `examples' of desired human-AI interactions through low-fidelity artifacts such as storyboards, prototype interfaces for task workflows, spreadsheets with ground truth data, and even interaction logs from existing non-AI software use. These artifacts are often ad hoc inventions intended to communicate with engineers about needed AI behavior. Third, given the challenges in articulating and reporting feedback about AI from end-users, designers share raw feedback from user testing through videos and direct observational notes and invite engineers to participate in end-user evaluation sessions. These new collaborative practices characterize the nature of leaky abstractions about designing AI components with end-users in mind. Finally, designers also offered technical representations such as qualitative code-books and epistemic design objects (including storyboards and prototypes) as shared representations for AI and UX specifications. Through these leaky abstractions, designers cross design-engineering boundaries to provide input about model behavior and training data. These design artifacts help engineers situate AI decisions within the broader context of human needs in HAI design. 

\begin{figure*}[hbtp]
\centering
\includegraphics[width=\textwidth]{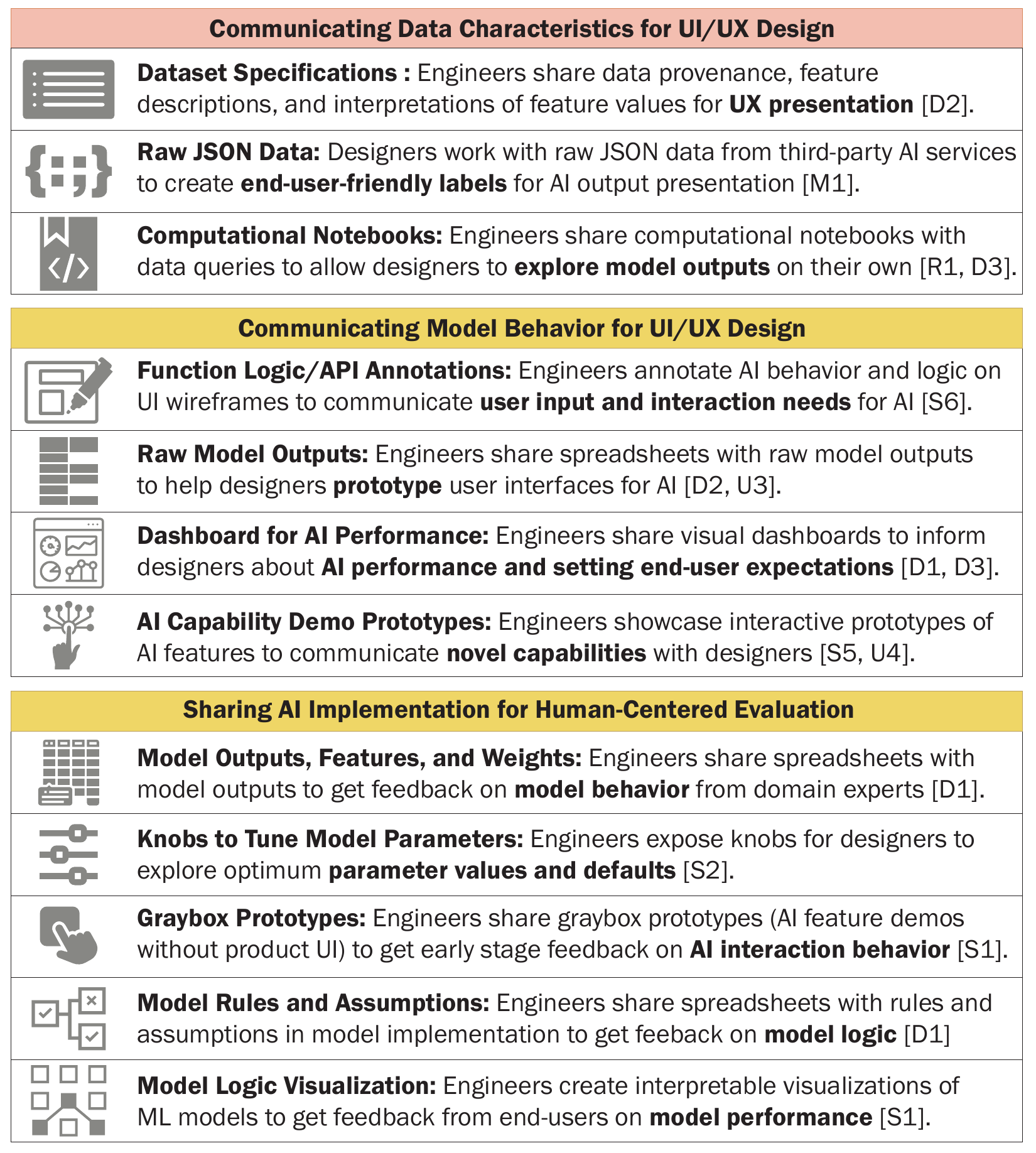}
\caption{Engineers Share AI Implementation Details with Designers to Inform UX Decisions}
\Description{Table of leaky artifacts shared by AI engineers with UX designers for communicating data characteristics for UI design, model behavior for UI design, and sharing implementation details for end-user evaluation.}
\label{fig:edleaks}
\end{figure*}

Similarly, engineers reported numerous leaky abstractions and novel collaboration practices to uncover AI implementation details for designers. As shown in Figure~\ref{fig:edleaks}, leaky abstractions allow engineers to (1) communicate about needed training data characteristics for user interface design, (2) communicate model behavior for user experience design, and (3) evaluate the AI with end-users. For instance, engineers assisted designers in exploring training data characteristics by creating and sharing computational notebooks with ready-to-run data queries and data specification documents. Access to these details supports designers in determining appropriate interface controls and presentation features, such as formatting and categorizing AI outputs. When prototyping ML models, engineers create envisioning prototypes to demonstrate capabilities and potential uses to designers. In other cases, they work with design teams to `align' model logic with interface designs by directly annotating over UI wireframes. Lastly, leaky abstractions showing AI logic, including interpretable visualizations, spreadsheets with model rules, and controls for tuning model parameters, allow designers to validate AI behavior with end-users and provide feedback on detailed AI implementations. 

\revisions{Each instance of leaky abstractions in the Figures~\ref{fig:deleaks} and \ref{fig:edleaks} may be difficult to anticipate, and may not be needed in a different project addressing a similar issue. This raises the question of whether standardized abstraction tools may be helpful, or if support is more helpfully aimed at training practitioners to invent their own leaky abstractions as needed. While a ``ready repository''~\cite{yang2018investigating} of abstractions  may solve a number of HAI challenges, useful generalizations and standard practices for abstractions may require more time to mature. This is especially true given the dynamic nature of AI tasks. In this regard, our definition of leaky abstractions is ingrained more specifically in software development and aimed toward shifting practitioners' mindsets towards integration rather than separation of concerns. Leaky abstractions appear useful in engaging designers in thinking about technical requirements, and in drawing engineers into thinking about how and what types of human data may improve system performance. In this sense, leaky abstractions may serve as a ``lingua franca'' to allow mutual consideration and problem solving to fuse human needs and AI capabilities in a design. From the findings collected, it is unclear whether there is substantial overlap in the problems arising and in the utility of specific leaky abstractions. Over the longer term, it is possible that a consensual set of useful representations and training may be helpful in supporting practitioners in this process.}

\revisions{Finally, leaky abstractions may appear to be similar to ``boundary negotiation artifacts'' used to form collaborative practices in situations where teams lack well-established standards~\cite{lee2007boundary}. In HAI, leaky artifacts can be seen as functioning in a similar manner by allowing designers and engineers to mutually alter the \textit{implementation} hierarchy---covering the product's function, specific implementation logic, and aggregation (part-whole) hierarchy---by representing how each component fits within the design. However, leaky abstractions differ from boundary negotiation artifacts in several important ways. First, they appear in our findings within collaborations where a well-established standard of practice (the conventional software SoC) already exists. There is no need to create boundaries because they are already well-known and practiced in software design; instead, leaky abstractions emerge when the established boundaries fail to support the needs of the design task. Second, the leaky artifacts observed emerge in response to a question or problem within a given design task; that is, the ad hoc nature of the representations suggest they are perceived to be useful in a specific collaboration and stage in the process. Finally, in a boundary negotiation, the desired result is a specification of how a separation of concerns is to be effected; in leaky abstractions, the impact is to facilitate sharing across concerns to collaborate on designs. Rather than form a new or different boundary, the outcome of using leaky abstractions is to ``break through'' a boundary within a circumscribed window of operation. The leaky abstraction provides a point for interchange across expertise situated within the present design task. Consequently, we conclude that leaky abstractions serve to enable specific collaborations about design decisions, and not to renegotiate boundary responsibilities in the design process. }

\subsection{Delayed Specifications Reduce Friction \revisions{in HAI Software Workflows}}
In conventional software workflows, best practice advice is to hide unimportant (and potentially complex) implementation detail across software modules and expertise boundaries~\cite{ousterhout2018philosophy}. In fact, any ``information leak'' about implementation is considered a `red flag'~\cite{ousterhout2018philosophy,helms2017leaky}. In this regard, our argument that leaky abstractions are \textit{necessary} for HAI development may seem counter-intuitive. However, our intent is not to argue against the power of  abstractions and established software practices. Rather, we suggest that abstraction may interfere when collaboration requires combining expertise. \revisions{Conventional software design may be best accomplished with a ``divide and conquer'' process; but in cases where integrative expertise is needed, collaboration may require sharing key lower level details (and not all). Because HAI development now requires fusing human needs within technical designs, points of intense collaboration across expertise roles will occur. Our findings document leaky abstractions as effective means for experts to jointly consider novel issues arising in HAI design.} 

\begin{figure*}[t!]
\centering
\includegraphics[width=\textwidth]{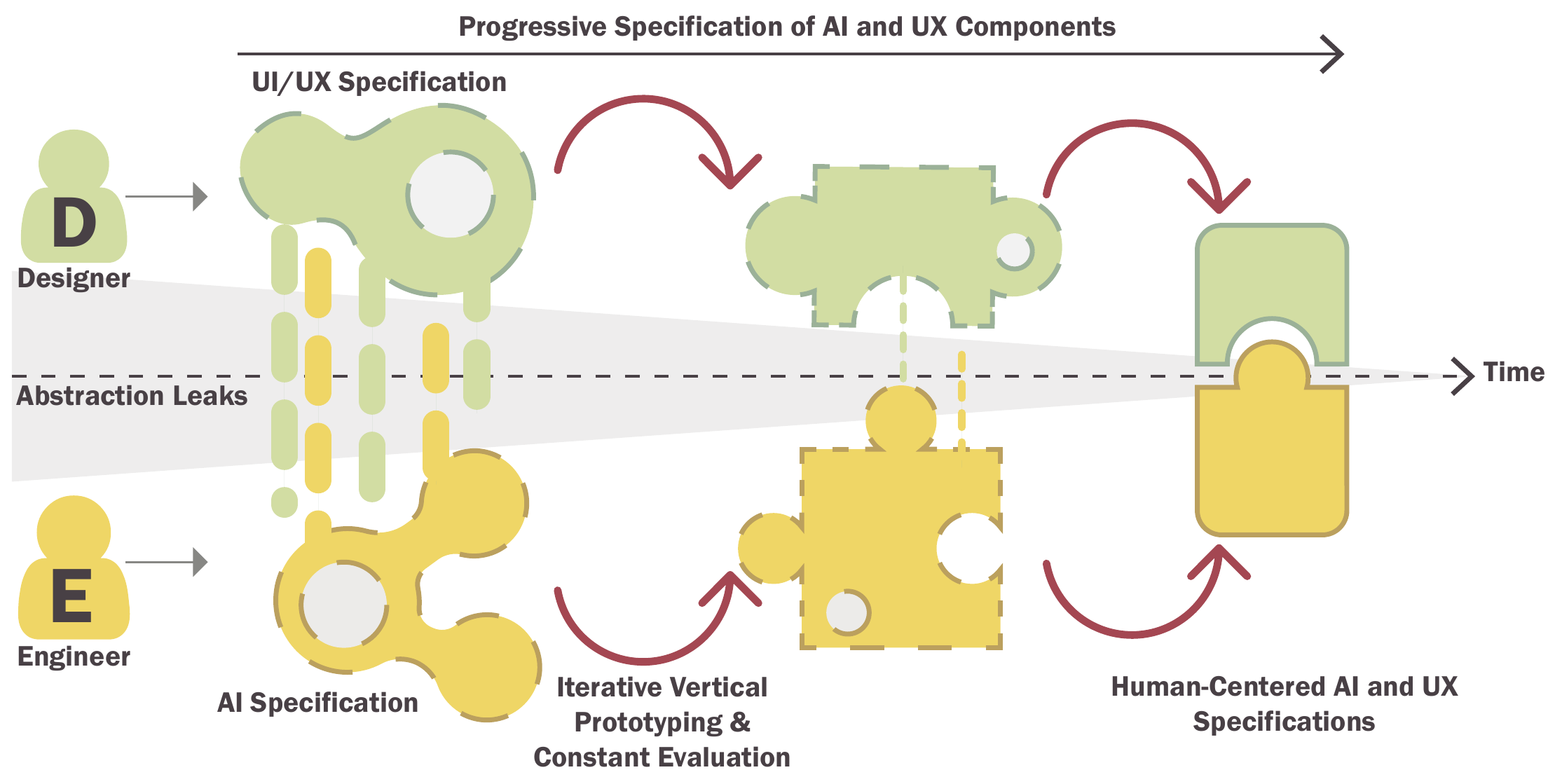}
\caption{Delayed Specification through Vertical Prototyping and Constant Evaluation}
\Description{High-level workflow for iterative prototyping AI and UX in parallel using leaky artifacts and constant evaluation}
\label{fig:deferred}
\end{figure*}

As shown in Figure~\ref{fig:deferred}, successful teams \textit{delayed} system specifications through iterative prototyping and constant evaluation. In the early design stages, designers and engineers produce fuzzy design specifications with some aspects more concretely defined. By sharing those initial design artifacts including low level details, teams overcome knowledge blindness to align AI and UX, and then collaboratively assess, negotiate, and revise their design choices. For instance, by sharing emerging AI behavior specifications, designers can evaluate assumptions and fit for end-users, update their own design representations for task workflows and interactions, and provide feedback for human-centered design of AI. During this stage, avoiding commitment to technical specifications affords later changes and invites collaboration and inputs. As the design progresses, more and more aspects of AI and UX components become concrete, and consequently, the need for leaky abstractions lessens. In the final design stages, successful teams arrive at realized designs \textit{solutions} aligned across AI, UX, and human users.

\revisions{Such a workflow could address critical concerns around responsible AI design. While not explicitly documented in this study, we suspect that the collaborative approach we identified may be helpful in anticipating problems in fairness, accessibility, and trustworthiness. For instance, SoC abstractions may lose key information about complex socio-technical contexts necessary for fairness~\cite{selbst2019fairabstract,agarwal2018reductions}. Selbst and colleagues argue that social context information may be critical for some design considerations, so standard abstraction methods may require alteration for HAI design. Similarly, our findings show that strict abstraction in representations shared between designers and engineers limit true collaboration. For instance, to design AI systems with fairness in mind, teams need to collaboratively define fair performance by considering diverse stakeholders, contexts of use, and assessment criteria (i.e., disaggregated evaluation~\cite{barocas2021designing}).  In developing their fairness checklist for software teams, Madaio et al. have enumerated steps for practitioners that span across all stages of the software lifecycle from product envisioning to deployment and maintenance~\cite{madaio2020co}. Further, teams need to anticipate how their design practices may limit considering diverse users. We imagine following the practice of delaying specification allows later changes and greater opportunities for responding to emerging information about system bias. Correspondingly, by sharing machine learning performance metrics and results (e.g., word error rate) with designers, engineers can better align model-level performance with diverse user needs and use contexts. In fact, our further research investigates collaborative practices for assessing fairness in AI systems~\cite{madaio2021assessing}. Despite increasing awareness, organizational goals and resource constraints continue to pose challenges for building responsible AI guidelines into current design practices~\cite{holstein2019improving,madaio2020co}. }

\subsubsection{Advice and Open Questions for Software Organizations:} Our findings on leaky abstractions and delayed specifications pose challenges for organizations creating AI applications. Typically, AI engineers and designers are situated in separate reporting structures (including different physical environments, incentives and performance reviews), and professional exchanges occur less often between professional groups. However, when designers and engineers successfully collaborate, the resulting human-AI system design succeeds. Their individual expertise requires that designers and engineers take very different perspectives on the design of AI applications. Engineers have a more immediate perspective of being ``in the weeds'' (M2), accountable for building specific functionality and ensuring the computational model is accurate and robust. Designers must take a future perspective by envisioning how an AI system might work, what human users might want, and how these work together seamlessly. HAI collaboration requires mutual respect for the expertise unique to each. As identified in our findings, leaky abstractions provide glimpses of the ``other side,'' windows large enough to help but not overwhelm or take over their own perspectives. In this dialogue across expertise, other team members may assist in translation. For example, project managers with a more holistic perspective may help facilitate cross-talk between engineers and designers. 

What might organizations do to promote the integrative collaboration practices identified in our findings? One suggestion is to increase points of contact for engineers and designers through co-working sessions. Regular informal interaction may help to identify intermediate points during the development process where integration is most needed. For instance, Subramonyam et al. propose a HAI design process model for early-stage co-design~\cite{subramonyamProcessModel2021}. This process may aid in catching emerging needs for hidden information in a just-in-time format rather than  at its end. In addition, regular discussions about data---so central to defining AI capabilities---may be helpful to both groups even without specific review goals. ``Data dives'' might share current observations, consider what data might better  inform choices, and review what is known about what users want. Providing space (in schedules and location) to inhabit the co-design process and build team familiarity will likely increase communication. Further, enacting the expectation that designers and engineers must share specifics about their progress facilitates the team co-design process within organizations.

\revisions{Further, to support co-design practices, as indicated by Yang et al.~\cite{yang2018investigating} documentation, development, and regularization of formats for leaky abstractions may be helpful. Organizations might build these formats based on current team experiences where shared leaky abstractions have proven beneficial.} When and why might a similar artifact be valuable in other design issues and projects? For example, engineers may create and collect leaky abstractions aimed at illustrating an AI's dynamic behaviors, such as showing how performance will change with more user data and demonstrating how failure cases arise. Designers may create and collect leaky abstractions to help engineers (and users) envision how a final application may feel to use, and how design choices may differently impact different users. As one engineering manager said, designers can \textit{``\ldots give people a target, and really paint a picture of that through design visualization, whether that's making a movie or just making a series of mocks, or building an experiential prototype, or something like that, really help land the tangibility of something pretty deep and complex''} (M2). How might designers create visualizations of user experiences to help engineers appreciate detailed information about the user's AI experience?  

Finally, the methods and tools available to designers require further innovation and development to respond to the need for considering more, and more varied, forms of data within the user experience. From simply testing prototypes with users, designers now need to consider data qualities representing what users want, how users  differ, who compiled databases represent, system data collected over users, users' personal data histories; and user feedback to systems. HAI systems can make use of much more information from human users to improve performance; this data is not at all abstract, but contained in individual examples. Off-the-shelf ML systems and data generation tools can support designers as they investigate alternative designs for data-intensive AI. New tools are being created; for example, Proto-AI is a prototyping tool for designers that can directly invoke AI models and services, incorporate model outputs into interface designs, and enable iterative and rapid evaluation of design choices across diverse end-users and data contexts ~\cite{subramonyam2021protoai}. However, additional robust methods and tools are needed to support designers in understanding the impact of data and providing targeted leaky abstractions for collaboration with engineers on HAI systems. 

\subsubsection{Advice for HCI and Software Pedagogy:} Ideally, to reduce the knowledge blindness identified in this work, HAI practitioners benefit from $\pi$-shaped expertise across HCI and AI (i.e., in-depth understanding of HCI \textit{and} AI)~\cite{baumer2017toward}. Quickly acquiring such knowledge is impractical given the rapidly advancing state-of-the-art in AI technology. However, as suggested above, it is essential to rethink the existing  emphasis on abstractions and separation of concerns in software development pedagogy. For AI application development, intensive data use requires fusing concerns across expertise roles. For instance, new software engineering courses might emphasize the importance of using leaky abstractions to communicate with UX professionals when developing AI-driven systems. Further, HCI pedagogy should equip future UX practitioners with data-driven design tools and methods to facilitate co-design. For instance, designers should receive training in incorporating data into their design and working with representations (e.g., interpretable ML) that occur at the boundaries. New toolkits and instructions can make HAI design accessible for students from differing backgrounds through supportive pedagogy and tools. Similarly, AI engineers should receive training in the parallel processes taking place in technical AI and UX design. They should be trained to understand the importance of UX in AI development, create and share representations of AI behavior before implementation, and co-design AI working across boundaries. Finally, the HAI curriculum should bring together students from varying backgrounds to engage in learning about team co-design throughout the engineering pipeline. Multidisciplinary pedagogical initiatives are essential to shaping the future of HAI as it is practiced in industry. 
    
\subsection{Limitations and Future work}
Our interviews with HAI practitioners provide insightful findings on boundary tensions between designers and engineers and their innovation in creating workarounds to share design ideas across boundaries. However, interview data may be limited compared to direct observation of work practices during \textit{in-situ} interactions. Due to non-disclosure agreements, our participants could not share some specific details about their HAI collaborations. For instance, interview participants were not able to share some representational artifacts they described. Consequently, our findings may fail to capture domain- and data-related nuances in conceptualizing leaky abstractions. \revisions{Further, our observations do not reveal factors related to team dynamics, power relationships, and company cultures.} Further research is needed to  investigate the distinct characteristics leaky abstractions used across  product domains, and practitioners' needs for techniques and tools to support them. Second, an essential aspect of HAI design is to meet responsible AI criteria. While our findings address the knowledge sharing and co-design practices necessary for various responsible AI criteria, future research should examine socio-technical practices and develop end-to-end strategies for enabling responsible AI. For instance, co-design will require innovative boundary artifacts and `leaks' to bridge fair performance metrics across AI and UX. Lastly, our work studies the boundaries between UX designers and AI practitioners. Future work should investigate how new emerging roles in HAI (such as ML-Operations practitioners), role-specific guidelines, and improved incentive structures, can provide organizational support for responsible AI design. 
\section{Conclusion}
In conventional software design, a clean separation of concerns between UX design and software implementation provides effective coordination and hand-off between designers and engineers on a team. However, there is no clean way to separate concerns when designing AI applications for human users. Instead, HAI demands points of greater integration in AI and UX design in order to address the burgeoning use of system dynamics and human data.  While our analysis of proposed HAI design guidelines in the field (and the HAI component model we construct from them) emphasized the necessity of multidisciplinary collaboration, little is known about how HAI systems are currently designed and developed in industry practice. Based on our interviews with UX researchers, AI engineers, data scientists, and project managers working on HAI applications, we identified current challenges in HAI design. Boundaries between designers and engineers introduce knowledge blindness about end-users and technology. For example, designers may not know the possibilities and limits of AI or be equipped to design for AI uncertainties. Engineers also describe difficulties in aligning data and AI models with end-user needs in the presence of uncertainty. Further, AI technology is based on a data-intensive approach that challenges conventional UX design practices. As a solution, we find that sharing leaky abstractions allows designers and engineers to overcome knowledge blindness and engage in collaborative HAI design. We offer an approach to collaboration that involves deferred specifications through iterative design and constant evaluation. Finally, we make recommendations for practice and pedagogy to support the collaborative creation of human-AI applications. 

\begin{acks}
We thank our reviewers and interview participants for their time and inputs. We also thank Steve Oney, Steven Drucker, Jasmine Jones, Tawfiq Ammari, and Yixin Zou for providing feedback on the paper. 
\end{acks}

\balance{}

 \bibliographystyle{ACM-Reference-Format}
 \bibliography{99_refs}

\end{document}